\documentclass[sn-basic,iicol]{sn-jnl}

\usepackage{graphicx}%
\usepackage{multirow}%
\usepackage{amsmath,amssymb,amsfonts}%
\usepackage{amsthm}%
\usepackage{mathrsfs}%
\usepackage[title]{appendix}%
\usepackage{xcolor}%
\usepackage{textcomp}%
\usepackage{subfigure}
\usepackage{manyfoot}%
\usepackage{booktabs}%
\usepackage{enumerate}
\usepackage{algorithm}%
\usepackage{algorithmicx}%
\usepackage{algpseudocode}%
\usepackage{listings}
\numberwithin{equation}{section}

\newcommand{\bx}{\mathbf{x}}
\newcommand{\bX}{\mathbf{X}}

\newcommand{\bbeta}{\boldsymbol{\beta}}

\theoremstyle{thmstyleone}%

\theoremstyle{thmstyletwo}%

\theoremstyle{thmstylethree}%

\begin{document}

\title{Online Prediction of Extreme Conditional Quantiles via B-Spline Interpolation}


\author[1]{\fnm{Zhengpin} \sur{Li}}\email{lizp21@m.fudan.edu.cn}

\author[1]{\fnm{Jian} \sur{Wang}}\email{jian\_wang@fudan.edu.cn}

\author*[1]{\fnm{Yanxi} \sur{Hou}}\email{yxhou@fudan.edu.cn}

\affil[1]{\orgdiv{School of Data Science}, \orgname{Fudan University}, \orgaddress{\street{Yangpu District}, \city{Shanghai}, \country{China}}}


\abstract{Extreme quantiles are critical for understanding the behavior of data in the tail region of a distribution. It is challenging to estimate extreme quantiles, particularly when dealing with limited data in the tail. In such cases, extreme value theory offers a solution by approximating the tail distribution using the Generalized Pareto Distribution (GPD). This allows for the extrapolation beyond the range of observed data, making it a valuable tool for various applications. However, when it comes to conditional cases, where estimation relies on covariates, existing methods may require computationally expensive GPD fitting for different observations. This computational burden becomes even more problematic as the volume of observations increases, sometimes approaching infinity. To address this issue, we propose an interpolation-based algorithm named EMI. EMI facilitates the online prediction of extreme conditional quantiles with finite offline observations.  Combining quantile regression and GPD-based extrapolation, EMI formulates as a bilevel programming problem, efficiently solvable using classic optimization methods. Once estimates for offline observations are obtained,  EMI employs B-spline interpolation for covariate-dependent variables, enabling estimation for online observations with finite GPD fitting. Simulations and real data analysis demonstrate the effectiveness of EMI across various scenarios.}


\maketitle

\section{Introduction}
In fields of statistical analysis such as risk management, finance, and environmental analysis, quantiles play a fundamental role in understanding data distributions and their various characteristics~(\citeauthor{allen2012does}, \citeyear{allen2012does}). Quantiles act as dividing points, segmenting the range of a probability distribution into continuous intervals with equal probabilities. Among the commonly used quantiles, the extreme quantile assumes a critical role, especially when rare events carry significant implications. This particular quantile sheds light on the tail behavior of a distribution and finds wide applications in decision-making, policy formulation, and strategic planning~(\citeauthor{resnick2008extreme}, \citeyear{resnick2008extreme}). Estimating the extreme quantile of a distribution, given a dataset comprising $N$ identically distributed observations, naturally involves using the sample quantile as a non-parametric estimation. This non-parametric approach performs admirably when the extreme quantile $\tau_N$ remains moderately high, i.e., as $\tau_N\rightarrow 1$ and $N(1-\tau_N)\rightarrow \infty$. In such cases, a sufficient number of observations lie above the quantile level $\tau_N$. However, when the observation size is not large enough and the observations above the more extreme quantile $\tau_N$ become scarce as $N(1-\tau_N)\rightarrow c,c\in[0,\infty)$, the non-parametric estimators are no longer inefficient and may cause serious underestimation~(\citeauthor{he2022risk}, \citeyear{he2022risk}). This presents a concern for accurately assessing risk, making informed financial decisions, and devising effective environmental strategies.

An alternative way to estimate extreme quantiles involves extrapolation beyond the range of observed data. {\bf E}xtreme {\bf V}alue {\bf T}heory (EVT) offers a systematic framework for precisely this purpose, allowing us to analyze the tail of a distribution and model the excess distribution above a prefixed threshold. The core principle of EVT is embodied in the famous Fisher-Tippett-Gnedenko theorem~(\citeauthor{balkema1974residual}, \citeyear{balkema1974residual}), which states that under certain conditions, the limiting distribution of extreme values can fall into the {\bf G}eneralized {\bf P}areto {\bf D}istribution (GPD). The excess distribution, denoted as $F_u(x)$, is defined as follows:
\begin{equation*}
	F_u(x) = \mathbb{P}(X-u \leq x \mid X>u), \quad 0\leq x<x_F-u,
\end{equation*}
where $x_F$ is the right endpoint of the distribution function $F(x)$, and $u$ serves as a large threshold. According to the Fisher-Tippett-Gnedenko theorem, as $u$ approaches the right endpoint, i.e., $u\rightarrow x_{F}$, the excess distribution converges to the GPD as:
\begin{equation*}
	\lim_{u\rightarrow x_{F}} F_u(x) = G_{\gamma,\sigma}(x),\quad 0\leq x<x_F-u,
\end{equation*}
where $G_{\gamma,\sigma}(x) = 1-(1+\gamma x / \sigma)_{+}^{-1 / \gamma}$ is the cumulative distribution function of the GPD. Here, \(\gamma \in \mathbb{R}\) and \(\sigma>0\) are called the shape and scale parameters, respectively. Various methods, such as the Pickands estimator~(\citeauthor{pickands1975statistical}, \citeyear{pickands1975statistical},) probability-weighted moments method~(\citeauthor{hosking1985estimation}, \citeyear{hosking1985estimation}), and maximum likelihood method~(\citeauthor{smith1987estimating}, \citeyear{smith1987estimating}), are available for estimating these parameters.

In light of the limiting distribution of extreme values, extensive research efforts have been dedicated to the estimation of extreme quantiles through extrapolation techniques. Among these methods, the {\bf B}lock {\bf M}axima {\bf M}ethod (BMM) stands out as a widely used approach~(\citeauthor{naveau2009modelling}, \citeyear{naveau2009modelling}). BMM involves partitioning the dataset into non-overlapping blocks and selecting the maximum observation within each block. The distribution of these block maxima is then fitted to the GPD to estimate extreme quantiles. A more general and flexible variant of BMM, known as the {\bf P}eak {\bf O}ver {\bf T}hreshold (POT) method, has gained more attention~(\citeauthor{ferreira2015block}, \citeyear{ferreira2015block}). Unlike BMM, which only considers the maximum observation in each block, POT broadens its scope to encompass several large values. To facilitate this, POT employs a threshold that isolates the exceedances, ensuring that they fall within the tail of the distribution. Subsequently, POT models these exceedances using the GPD, allowing for the estimation of extreme quantiles with increased flexibility and accuracy.

While the BMM and POT have proven effective in estimating extreme quantiles, they exhibit a limitation. In essence, they do not account for dependencies between variables. Consequently, these methods may not be suitable for scenarios where extreme events are influenced by complex interplays of covariates. To address this limitation, conditional quantile regression emerges as a powerful extension of classical regression, offering a comprehensive approach to modeling the conditional quantiles of covariates~(\citeauthor{koenker2001quantile}, \citeyear{koenker2001quantile}). Consider a two-variable case where the distribution of a response variable $Y$ depends on a set of covariates $\mathbf{X} \in \mathbb{R}^p$. Here, $p$ denotes the feature dimension. In this context, the conditional quantile is defined as $Q_Y(\tau|\bx)=F_{Y}^{-1}(\tau | \mathbf{x})$, where $F_{Y}^{-1}(\cdot | \mathbf{x})$ is the generalized inverse of the conditional distribution function of $Y$. When the extreme quantile $\tau$ is moderately high, $Q_{Y}(\tau |\bx )$ can be efficiently estimated via simple models like linear conditional quantile regression. However, when dealing with scenarios where the quantile $\tau = \tau_N$ satisfies 
\begin{equation*}
	N(1-\tau_N)\rightarrow c,c\in[0,\infty),
\end{equation*}
these estimation methods may introduce substantial biases.

In such cases, the combination of EVT and conditional quantile regression becomes essential. One common approach relies on the fact that, under mild conditions,
\begin{equation*}
	\lim_{t\rightarrow\infty}\frac{U(tz | \bx) - U(t | \bx)}{a(t)} = \frac{z^{\gamma} - 1}{\gamma},\quad z>0,
\end{equation*}
where $\gamma$ is a real parameter, $U(\cdot|\bx)$ is the inverse function of $1/1-F(\cdot|\bx)$ and $a(\cdot)$ is a suitable positive function.
Please see~(\citeauthor{li2019extreme}, \citeyear{li2019extreme}),  (\citeauthor{wang2013estimation}, \citeyear{wang2013estimation}), (\citeauthor{wang2012estimation}, \citeyear{wang2012estimation}),  (\citeauthor{xu2022prediction}, \citeyear{xu2022prediction}) for more details. In this paper, we focus on an alternative way that employs the GPD to approximate the tail distribution. The resulting methods for extreme conditional quantile regression typically involve two key components: 
i) first estimating the intermediate conditional quantile $Q_Y(\tau_0|\bx)$ at an intermediate level $\tau = \tau_0$ through classic conditional quantile regression, and ii) then modeling the exceeding data above the threshold $u = \widehat{Q}_Y(\tau_0|\bx)$ using the GPD. Notable examples of such methods can be found in the work of~\citeauthor{hou2022three} (\citeyear{hou2022three}).

While existing extreme conditional quantile regression methods have achieved significant success, they exhibit certain limitations. These limitations become apparent when the response variable $Y$ is influenced by covariates $\mathbf{X}$, as the conditional quantile $Q_Y(\tau_0|\bx)$ becomes dependent on $\mathbf{X}$. In other words, the threshold $u = {Q}_Y(\tau_0|\bx)$ varies with different observations $\mathbf{x}$, resulting in different sets of exceeding data used to fit the GPD. Consequently, both the scale and shape parameters of the GPD also become dependent on $\mathbf{X}$. To estimate these covariate-dependent parameters, one must conduct conditional quantile regression and extreme distribution fitting once for each observation. This is computationally complex since most existing methods are time-consuming in the above two steps~(\citeauthor{velthoen2019improving}, \citeyear{velthoen2019improving}). More critically, practical scenarios often involve online streaming data, where the observations continually emerge over time. In such cases, the dataset's size can become substantial or even infinite~(\citeauthor{granello2004online}, \citeyear{granello2004online}). Under this circumstance, applying existing methods becomes impractical, as performing GPD fitting for each observation becomes computationally burdensome and unacceptable.

To solve this issue, in this paper, we propose an {\em offline} algorithm denoted as EMI (stands for {\bf E}xtrapolation via {\bf M}athematical programs with equilibrium constraints and B-spline {\bf I}nterpolation). Given finite offline historical observations, our algorithm contains two main stages. First, EMI employs linear conditional quantile regression to estimate an intermediate quantile, ${Q}_Y(\tau_0|\bx)$, which serves as the threshold. The exceedances beyond this threshold are modeled using a GPD with scale and shape parameters that depend on the covariates. Notably, EMI innovatively formulates these regression and fitting tasks as a bilevel programming problem, where the lower- and upper-level functions represent the check function and maximum likelihood function, respectively. For each offline observation, we obtain the corresponding estimated parameters by solving the bilevel programming with classic optimization methods, such as interior point algorithms and active set algorithms. Second, EMI employs B-spline interpolation for the scale and shape parameters. The interpolation is conducted on each feature dimension of the covariates. For online observations with features falling outside the range of offline data, EMI leverages the nearest available data points for estimation. This innovative approach eliminates the need for infinite repeated GPD fitting when encountering new online data. Numerical simulations and real-world analysis further illustrate the effectiveness of EMI in extreme conditional quantile estimation.

\section{Methodology}
\subsection{Conditional Quantile Estimation via GPD Approximation}
Conditional quantile estimation is a statistical technique that focuses on modeling and analyzing conditional quantiles of a response variable concerning one or more predictor variables. The techniques involved in conditional quantile estimation include quantile regression and various non-parametric approaches. The motivation behind conditional quantile estimation is typically different from that of traditional mean-based modeling. While traditional regression techniques, like linear regression, focus on estimating the conditional mean of the response variable, conditional quantile methods seek to estimate quantiles, which offer a more comprehensive view of the conditional distribution. This approach is particularly useful when dealing with non-normally distributed data, outliers, or when the interest is in understanding the variability at different points in the distribution.

We consider here the setting where the response $Y \in \mathbb{R}$ depends on covariates $\mathbf{X} \in \mathbb{R}^p$. Denote the conditional distribution of the variable $Y$ as $F_Y(\cdot|\bX)$ and marginal distribution of $Y$ as $F_Y$. Suppose observations $(y_i,\bx_i),i=1,2,\ldots,N$, where $y_i$ is generated from $F_Y(\cdot|\bx_i)$ given $\bX=\bx_i$. The conditional quantile of $Y$ given $\bX=\bx$ at the quantile level $\tau\in(0,1)$ is defined as $Q_Y(\tau|\bx)=F_{Y}^{-1}(\tau | \mathbf{x})$, where $F_{Y}^{-1}(\cdot | \mathbf{x})$ is the generalized inverse of the conditional distribution function of $Y$. The conditional quantile $Q_Y(\tau|\bx)$ can be calculated as the solution to the following optimization problem 
\begin{equation}\label{eq:quantile_E}
	Q_Y(\tau|\bx) = \arg \min _{q \in \mathbb{R}} \mathbb{E}\left[\rho_{\tau}(Y-q)| \bX = \bx\right],
\end{equation}
where $\rho_\tau(u)=(\tau-I(u\le 0))u$ is the so-called check function~(\citeauthor{koenker1978regression}, \citeyear{koenker1978regression}). Quantile regression is one of the most widely used techniques for conditional quantile estimation. It extends traditional regression by estimating the conditional quantiles directly. We follow the standard parametric assumption that the true conditional quantile has a linear form as
\begin{equation}\label{def:qr}
	Q_Y(\tau|\bx)=\alpha(\tau)+\bbeta^T(\tau)\bx,
\end{equation}
where $(\alpha(\tau),\bbeta^T(\tau))$ is a parameter vector associated with the  level $\tau$ and $(\alpha_0(\tau),\bbeta_0^T(\tau))$ denotes the true parameter. With observations $(y_i,\bx_i),\,i=1,2,\ldots, N$, the parameter can be estimated by solving an optimization problem corresponding to the empirical counterpart of~\eqref{eq:quantile_E}:
\begin{equation}\label{est:qr}
	\begin{split}
		(\widehat{\alpha}(\tau), \widehat{\bbeta}^T(\tau)) &= \arg \min _{\alpha,\bbeta} L^{(\text{QR})}_{N\tau}(\alpha,\bbeta)\\ &= \arg \min _{\alpha,\bbeta} \sum_{i=1}^N\rho_{\tau}(y_i-\alpha-\bbeta^T\bx_i).
	\end{split}
\end{equation}
The conditional quantile estimation of $Y$ given $\bX=\bx$ is thus calculated as
\begin{equation}\label{eq:quantile_emp}
	\widehat{Q}_Y(\tau|\bx) = \widehat{\alpha}(\tau) + \widehat{\bbeta}^T(\tau)\bx.
\end{equation}

The empirical linear estimation of the conditional quantile works effectively when the quantile level $\tau = \tau_0$ is moderately high. However, as the quantile level $\tau = \tau_N$ approaches $1$ and $N(1-\tau_N)\rightarrow c,c\in[0,\infty)$, the expected number of observations exceeding $Q_Y(\tau_N|\bx)$ becomes insufficient. In such cases, empirical linear estimation of the conditional quantile is no longer feasible, necessitating extrapolation beyond observed data. Extrapolation is the process of estimating or predicting values beyond the range of the available data. It is a common practice in various fields, including finance, environmental science, insurance, and reliability analysis, where the primary interest lies in understanding and managing extreme events, such as financial market crashes, natural disasters, rare disease occurrences, or equipment failures. To address this need for extrapolation, we turn to the GPD, which offers an excellent approximation of the tail distribution of the conditional distribution of $Y$ given $\bX=\bx$. More specifically, given a large threshold $u$, $\bX = \bx$, and $z>0$, the exceedance \(Y-u \mid Y>u\) given $\bX=\bx$ satisfies
\begin{equation}\label{def:gpd}
	\begin{split}
		&P\left(Y > u +z\,|\,Y> u, \bX=\bx\right)\\
		=&\left\{\begin{matrix}\left(1+\frac{\gamma_0(\bx) z}{\sigma_0(\bx;\tau_u(\bx))}\right)^{-1/\gamma_0(\bx) }, & \gamma_0(\bx)\ne 0, \\ \exp\left(-\frac{z}{\sigma_0(\bx;\tau_u(\bx))}\right),&\gamma_0(\bx)=0, \end{matrix}\right.
	\end{split}
\end{equation}
where $\tau_u(\bx)=F_Y(u|\bx)\in(0,1)$ and we require $1+\gamma_0(\bx)z/\sigma_0(\bx;\tau_u(\bx)) >0$ for $\gamma_0(\bx)\neq 0$. Here $\gamma_0(\bx)$ is the extreme value index, and $\sigma_0(\bx;\tau_u(\bx))$ is a scale related to the quantile level of the threshold $u$. When \(\gamma_{0}(\bx)<0\), there is a finite right endpoint \(u^{*}=u-\frac{\sigma_0(\bx;\tau_u(\bx))}{\gamma_{0}(\bx)}\) in the support of the distribution of \(Y\), that is, \(F_Y(y|\bX)=1\) for all \(y \geq u^{*}\). When \(\gamma_{0}(\bx)=0\), the exceedance \(Y-u \mid Y>u\) given $\bX=\bx$ has an exponential distribution with mean \(\sigma_0(\bx;\tau_u(\bx))\). When \(\gamma_{0}(\bx)>0\), the exceedance \(Y-u \mid Y>u\) given $\bX=\bx$ has a heavy tail with up to \(\frac{1}{\gamma_{0}(\bx)}\)-th finite moments. Note that for any higher threshold $u'>u$, the exceedance \(Y-u \mid Y>u\) given $\bX=\bx$ again follows the generalized Pareto distribution with the same shape parameter $\gamma_0(\bx)$ but a different scale parameter.

Once the threshold $u$ is set, the next step involves estimating the parameters of the GPD. In the absence of covariates, a standard way of estimating the GPD parameters is the maximum likelihood method, which provides asymptotically normal estimators in the unconditional case with $\gamma > -1/2$~(\citeauthor{drees2004maximum}, \citeyear{drees2004maximum}). To estimate the covariate-dependent variables $\gamma_0(\bx)$ and $\sigma_0(\bx;\tau_u(\bx))$, we also use the maximum likelihood method. Specifically, given observations $(y_i,\bx_i),\,i=1,2,\ldots, N$ and the threshold $u$, we define the exceedances of the observations above the threshold as
\begin{equation}
	z_{i}=\left(y_{i}-u\right)_{+}, \quad i=1, \ldots, N,
\end{equation}
where $(a)_+ = \max\{0,a\}$. In other words,  \(z_{i}=0\) whenever the value \(y_{i}\) is below the threshold. Denoe the log-likelihood function as 
\begin{equation}\label{log-likelihood}\small
	\begin{split}
		l(\gamma,\sigma|z)=&-\left\{\frac{1+\gamma}{\gamma}\log\left(1+\frac{\gamma z}{\sigma}\right)+\log\sigma\right\}I(\gamma\ne 0)\\&-\left\{\frac{z}{\sigma}+\log\sigma\right\}I(\gamma=0),
	\end{split}
\end{equation}
and the loss function is calculated only on positive $z_i$ as 
\begin{equation*}
	L^{(\text{MLE})}_N(\gamma,\sigma|u)=-\sum_{i=1}^Nl(\gamma,\sigma|z_i)I(z_i > 0).
\end{equation*}
Then, to estimate $\gamma_0(\bx)$ and $\sigma_0(\bx;\tau_u(\bx))$, we can apply the censored maximum likelihood method via solving the following optimization problem
\begin{equation}\label{est:gpd}
	\max_{\gamma,\sigma} L^{(\text{MLE})}_{N}(\gamma,\sigma|u).
\end{equation}

Inverting the distribution function in~\eqref{def:gpd}, we have an approximation of the quantile for probability level $\tau_N > \tau_0$ by taking $u=Q_Y(\tau_0|\bx)$ and $z=Q_Y(\tau_N|\bx)-Q_Y(\tau_0|\bx)$. As a result, we have that
\begin{equation}\label{eq:gpd}\small
	\begin{split}
		&Q_Y(\tau_N|\bx)\\
		=&\left\{\begin{matrix}Q_Y(\tau_0|\bx)+\frac{\sigma_0(\bx;\tau_0)}{\gamma_0(\bx)}\left(\tau_{0,N}^{\gamma_0(\bx)}-1\right), & \gamma_0(\bx)\ne 0, \\ Q_Y(\tau_0|\bx)+\sigma_0(\bx;\tau_0)\log\tau_{0,N}, & \gamma_0(\bx)=0, \end{matrix}\right.
	\end{split}
\end{equation}
where for simplicity, we define
\begin{equation}\label{tau0N}
	\tau_{0,N} = \frac{1-\tau_0}{1-\tau_N}.
\end{equation}
We denote the solution of the problem~\eqref{est:gpd} as $\widehat\gamma(\bx)$ and $\widehat\sigma(\bx;\tau_0)$. Therefore, using plug-in estimators, the prediction of the extreme conditional quantile $Q_Y(\tau_N | \bx)$ is 
\begin{equation}\label{plug-in}\small
	\begin{split}
		& \widehat Q_Y(\tau_0|\bx)=\widehat\alpha(\tau_0)+\widehat\bbeta^T(\tau_0)\bx, \\
		& \widehat Q_Y(\tau_N|\bx)\\
		=&\left\{
		\begin{matrix}\widehat Q_Y(\tau_0|\bx)+\frac{\widehat\sigma(\bx;\tau_0)}{\widehat\gamma(\bx)}\left(\tau_{0,N}^{\widehat\gamma(\bx)}-1\right), & \widehat\gamma(\bx)\ne 0, \\ \widehat Q_Y(\tau_0|\bx)+\widehat\sigma(\bx;\tau_0)\log\tau_{0,N}, & \widehat\gamma(\bx)=0. 
		\end{matrix}\right.
	\end{split}
\end{equation}

It is important to note that the estimations in \eqref{est:qr} and \eqref{est:gpd} are usually implemented individually, such as the stages two and three of the algorithm in \cite{hou2022three}. However, the combination of \eqref{est:qr} and \eqref{est:gpd} is essentially a bilevel programming problem when the threshold $u$ in \eqref{est:gpd} equals the conditional quantile $Q(\tau|\bx)$ determined by \eqref{def:qr} and \eqref{est:qr}. In the following, we apply algorithms of bilevel programming to obtain the joint estimation in \eqref{est:qr} and \eqref{est:gpd}.

\subsection{Stage 1: Offline Quantile Estimation via Bilevel Programming}\label{sec2.2}
In this section, we formulate the extreme quantile estimation via GPD approximation as a bilevel programming problem. Bilevel programming, a class of optimization problems characterized by their hierarchical structure, has found wide-ranging applications across diverse domains including economics, engineering, transportation, and environmental management~(\citeauthor{sinha2017review}, \citeyear{sinha2017review}). These problems feature a unique interplay between two optimization tasks: an upper-level decision-maker seeks to optimize a given objective while factoring in the response of a lower-level decision-maker, who, in turn, optimizes their objective in response to the upper-level decisions. To illustrate this concept in our context, we combine~\eqref{est:qr} and~\eqref{est:gpd} and obtain
\begin{equation}\label{opt:bilevel}
	\begin{split}
		\max_{\gamma,\sigma}\quad & L^{(\text{MLE})}_{N}(\gamma,\sigma|Q_Y(\tau_0|\bx))\\
		 s.t.\quad & \min_{\alpha,\bbeta} L^{(\text{QR})}_{N\tau_0}(\alpha,\bbeta),
	\end{split}
\end{equation}
where $\bx$ and the prefixed level $\tau_0$ is given. The conditional quantile regression and the maximum likelihood function are denoted as the lower- and upper-level functions of \eqref{opt:bilevel}, respectively.

The hierarchical structure of bilevel programming introduces unique challenges that pose difficulties for existing solution methods. For instance, penalty function methods address bilevel optimization by solving a sequence of unconstrained optimization problems. These unconstrained problems are derived by introducing a penalty term that quantifies the extent of constraint violation. The penalty term typically relies on a parameter, taking a zero value for feasible solutions and a positive value (indicating minimization) for infeasible ones. However, many penalty function methods require the upper-level function to be convex, which conflicts with the non-convex nature of the log-likelihood function~\eqref{log-likelihood}. This discrepancy arises from the fact that the Hessian matrix $\frac{\partial l(\gamma,\sigma|z)}{\partial \gamma \partial \sigma}$ is not semi-positive definite. Consequently, various penalty methods, such as those proposed by \citeauthor{aiyoshi1981hierarchical} (\citeyear{aiyoshi1981hierarchical}) and \citeauthor{lv2007penalty} (\citeyear{lv2007penalty}), tend to be ineffective in addressing the problem. Additionally, nested evolutionary algorithms have gained popularity for dealing with bilevel problems (\citeauthor{sinha2014finding}, \citeyear{sinha2014finding}). These algorithms involve solving a lower-level optimization problem for each upper-level member, which are typically employed in two main ways. The first approach combines an evolutionary algorithm at the upper level with classical algorithms at the lower level (\citeauthor{mathieu1994genetic}, \citeyear{mathieu1994genetic}). In contrast, the second approach leverages evolutionary algorithms at both the upper and lower levels (\citeauthor{angelo2013differential}, \citeyear{angelo2013differential}). While these nested strategies can be effective, they face a significant computational burden, rendering them impractical for large-scale bilevel problems.

To handle problem~\eqref{opt:bilevel} effectively, we first reformulate the lower-level function as
\begin{equation*}
	\begin{split}
		\min _{\alpha, \boldsymbol{\beta}} & \sum_{y_i>\boldsymbol{\alpha}+\boldsymbol{\beta}^T \mathbf{x}_i} \tau_0\left(y_i-\alpha-\boldsymbol{\beta}^T \mathbf{x}_i\right)\\
		&+\sum_{y_i<\alpha+\boldsymbol{\beta}^T \mathbf{x}_i}(\tau_0-1)\left(y_i-\alpha-\boldsymbol{\beta}^T \mathbf{x}_i\right).
	\end{split}
\end{equation*}
The above problem can be transformed into a linear program as
\begin{equation}\label{LP}
	\begin{aligned}
\min _{\alpha, \boldsymbol{\beta}, \lambda_i^{+}, \lambda_i^{-}} & \sum_{i=1}^N \tau_0 \lambda_i^{+}+(1-\tau_0) \lambda_i^{-}, \\
{s.t. }\quad & y_i=\alpha+\boldsymbol{\beta}^T \mathbf{x}_i+\lambda_i^{+}-\lambda_i^{-}, i=1, \ldots, N, \\
& \lambda_i^{+}, \lambda_i^{-} \geq 0, i=1, \ldots, N,
\end{aligned}
\end{equation}
where $\lambda_i^+,\lambda_i^-,i = 1,2,\ldots,N$ are Lagrange multiplicators. With \eqref{LP}, the bilevel programming \eqref{opt:bilevel} can be rewritten as
\begin{equation}\small
	\begin{split}\label{opt:rebilevel}
	\max_{\alpha, \bbeta, \gamma, \sigma}&-\sum_{i=1}^Nl\left(\gamma,\sigma|y_i-\alpha - \bbeta^T \bx\right)I\left(y_i-\alpha-\bbeta^T \bx\right)\\
		{s.t.} \quad&
		\{\alpha, \bbeta, \lambda^+_i,\lambda^-_i\} \in \mathcal{A},\quad i=1,2,\ldots,N,
	\end{split}
\end{equation}
where we let $z_i = (y_i-u)_+$ and $u = \alpha+\bbeta^T \bx$. Then, we replace the constraints in problem~\eqref{opt:rebilevel} with its Karush-Kuhn-Tucker (KKT) conditions. These conditions manifest as Lagrangian and complementarity constraints, simplifying the bilevel optimization problem into a single-level constrained optimization problem. Define the augmented Lagrangian (AL) function as
\begin{equation*}
	\begin{split}
		L_N^{(\text{AL})} &= \sum_{i=1}^N~ \tau_0 \lambda_i^{+}+(1-\tau_0) \lambda_i^{-} \\
		&+ \sum_{i=1}^N u_i(y_i-\alpha-\boldsymbol{\beta}^T \mathbf{x}_i-\lambda_i^{+}+\lambda_i^{-}) \\
		&+ \sum_{i=1}^N t_i^+\lambda_i^{+}+t_i^-\lambda_i^{-},
	\end{split}
\end{equation*}
where $t_i^+,t_i^-,i = 1,2,\ldots,N,$ are the Lagrange multiplicators. The derivatives of $L_N^{(\text{AL})}$ with respect to the parameters $\alpha, \bbeta, \lambda^+_i,\lambda^-_i$ are given as follows:
\begin{equation}\small
	\begin{split}
		\nabla_\alpha L_N^{(\text{AL})} = 0 ~&~\Rightarrow \sum_{i=1}^{N} u_i = 0, \\
		\nabla_{\bbeta} L_N^{(\text{AL})} = 0 ~&~\Rightarrow \sum_{i=1}^{N} u_i\bx_i = 0,\\ 
		\nabla_{\lambda_i^{+}} L_N^{(\text{AL})} = 0 ~&~\Rightarrow \tau_0 - u_i - t^+_i = 0, \\
		\nabla_{\lambda_i^{-}} L_N^{(\text{AL})} = 0 ~&~\Rightarrow 1-\tau_0 + u_i - t^-_i = 0. \\
	\end{split}
\end{equation}
Therefore, the problem~\eqref{opt:rebilevel} is reformulated as
\begin{equation}\label{eq:MPEC}\small
	\begin{split}
	\max_{\Theta_{\bx,\tau_0}} ~&-\sum_{i=1}^N\ell\left(\gamma,\sigma|y_i-\alpha - \bbeta^T \bx\right)I\left(y_i-\alpha-\bbeta^T \bx\right)\\
	{s.t.} \quad &~	y_i = \alpha+\bbeta^T \bx+\lambda^+_i-\lambda^-_i, \quad i=1,\ldots,N, \\
		&~	t^+_i, t^-_i, \lambda^+_i,\lambda^-_i \geq 0, \quad i = 1,\ldots,N, \\
		&~	t^+_i\lambda^+_i = t^-_i\lambda^-_i = 0, \quad i = 1,\ldots,N, \\
		&~  \sum_{i=1}^{N} u_i = 0, \quad\sum_{i=1}^{N} u_i\bx(i) = 0, \\
		&~	\tau - u_i - t^+_i = \quad i = 1,\ldots,N, \\
		&~ 1-\tau + u_i - t^-_i = 0, \quad i = 1,\ldots,N,
	\end{split}
\end{equation}
where $\Theta_{\bx,\tau_0} = (\gamma, \sigma, \alpha, \bbeta, \lambda^+_i,\lambda^-_i, t^+_i, t^-_i, u_i)$ denotes the variable set for a prefixed $\bx$ and level $\tau_0$. Problem \eqref{eq:MPEC} is a special case of mathematical programs with equilibrium constraints (MPEC), in which the constraints include the complementarity conditions
\begin{equation*}
	t^+_i\lambda^+_i = t^-_i\lambda^-_i = 0, \quad i = 1,2,\ldots,N.
\end{equation*}
MPEC plays an important role in various fields, such as the Stackelberg game in economic sciences, and problem \eqref{eq:MPEC} could be efficiently solved via interior point algorithm or active set algorithm. For more details on these algorithms, we refer readers to the work of \cite{luo1996mathematical}. In the following, we leave out the solutions of the Lagrange multiplicators. and denote the solutions of problem \eqref{eq:MPEC} as $\Theta_{\bx,\tau_0} = (\widetilde{\alpha}(\tau_0), \widetilde{\bbeta}(\tau_0), \widetilde\gamma(\bx),\widetilde\sigma(\bx;\tau_0))$ for simplicity.

\subsection{Stage 2: Online Prediction via B-Spline Interpolation}
Note that the solution to \eqref{eq:MPEC} is derived given a specific covariate $\bx$. This leads to estimations $\widetilde\gamma(\bx)$ and $\widetilde\sigma(\bx;\tau_0)$ that are inherently covariate-dependent. To illustrate this, recall that $\widetilde\gamma(\bx)$ and $\widetilde\sigma(\bx;\tau_0)$ are utilized in modeling the exceedances over a threshold. We estimate this threshold through \eqref{eq:quantile_emp}, based on a prefixed $\tau_0$. Consequently, the estimated threshold becomes intricately linked to the covariates, $\bx$. This, in turn, results in varying exceedances for different values of $\bx$, ultimately leading to distinct GPD models and covariate-dependent estimations, namely $\widetilde\gamma(\bx)$ and $\widetilde\sigma(\bx;\tau_0)$.

In practice, we often encounter situations where we need to deal with an infinite number of future observations, in addition to the finite historical observations at hand. A typical example is the realm of online streaming data, continually generated from various sources~(\citeauthor{gaber2005mining}, \citeyear{gaber2005mining}). Online streaming data finds applications across diverse industries, including finance, transportation, and e-commerce. Notably, streaming data volumes can be vast, occasionally approaching infinity. When it comes to the task of conditional quantile estimation with these infinite observations from online streaming data, applying \eqref{eq:MPEC} entails solving the problem an infinite number of times. This introduces a significant computational complexity that poses a substantial challenge.

To alleviate the computational burden associated with solving \eqref{eq:MPEC} for possible infinite observations, we employ B-spline interpolation. To clarify, let us denote the finite offline observations as $(y_i^{(\text{off})},\bx_i^{(\text{off})})$ for $i=1,\ldots,N^{(\text{off})}$ and the future online observations as $\bx_i^{(\text{on})}$ for $i=1,\ldots,N^{(\text{on})}$ where we allow $N^{(\text{on})}$ to tend to infinite. It is important to note that we assume the covariates for the two observation sets are distinct. B-spline interpolation empowers us to solve \eqref{eq:MPEC} only $N^{(\text{off})}$ times, obtaining estimations for any number of data points, even as $N^{(\text{on})}$ approaches infinity. In more specific terms, assume that we have obtained $\Theta_{\bx_i^{(\text{off})}, \tau_0} = (\widetilde{\alpha}(\tau_0), \widetilde{\bbeta}(\tau_0), \widetilde\gamma(\bx_i^{(\text{off})}),\widetilde\sigma(\bx_i^{(\text{off})};\tau_0))$ for $i=1,\ldots,N^{(\text{off})}$, for given offline observations by solving \eqref{eq:MPEC}. Notably, the estimations $\widetilde{\alpha}(\tau_0)$ and $\widetilde{\bbeta}(\tau_0)$ are solely linked to the choice of $\tau_0$, which is assumed to be the same for both offline and online observations in our paper. As a result, for $\bx_i^{(\text{on})}$, the estimations of $\alpha$ and $\bbeta$ remain consistent with those for $\bx_i^{(\text{off})}$. Therefore, we only need to apply B-spline interpolation to estimate the extreme value index $\gamma(\bx)$ and the scale $\sigma(\bx,\tau_0)$ for $\bx = \bx_i^{(\text{on})}$.

B-splines, which stand for ``Basis splines'', hold a pivotal role in the realm of numerical analysis and computer-aided design. They provide a powerful method for representing and approximating complex curves and surfaces, as well as for solving various mathematical and engineering problems. 
B-splines are piecewise polynomials, and the positions where the pieces meet are known as knots, which are defined as a non-decreasing sequence of real numbers
\begin{equation*}
	\boldsymbol{\xi}=\left\{\xi_{i}\right\}_{i=1}^{M}=\left\{\xi_{1} \leq \xi_{2} \leq \cdots \leq \xi_{M}\right\},
\end{equation*}
where $M$ is the knot number. Provided that $M\geq d +2$, we can define the B-splines of degree $d\geq0$ over the knots $\xi$ using a recursive formulation. Specifically, the $j$-th B-spline $B_{j, d, \boldsymbol{\xi}}(x):\mathbb{R}\rightarrow \mathbb{R}$ can be calculated by
\begin{equation*}
	\begin{split}
		B_{j, d, \boldsymbol{\xi}}(x):=&\frac{x-\xi_{j}}{\xi_{j+d}-\xi_{j}} B_{j, d-1, \boldsymbol{\xi}}(x)\\
		&+\frac{\xi_{j+d+1}-x}{\xi_{j+d+1}-\xi_{j+1}} B_{j+1, d-1, \boldsymbol{\xi}}(x),
	\end{split}
\end{equation*}
where \(\xi_{j} \leq \xi_{j+1} \leq \cdots \leq \xi_{j+d+1}\) are \(d+2\) real numbers taken from the knot sequence \(\xi\). When $\xi_{j}=\xi_{j+d+1}$, $B_{j, d, \boldsymbol{\xi}}(x)$ is identically zero. The foundation of B-splines begins with the initial B-spline of order zero, represented as:
\begin{equation*}
	B_{i, 0, \boldsymbol{\xi}}(x):=\left\{\begin{array}{ll}1, & \text { if } x \in\left[\xi_{i}, \xi_{i+1}\right), \\ 0, & \text { otherwise }.\end{array}\right.
\end{equation*}
In simpler terms, a B-spline operates over $M$ knots, strictly adhering to their non-decreasing order. B-splines contribute meaningfully only within the range defined by the first and last knots, remaining zero elsewhere. It is worth noting that while two B-splines can share some of their knots, two B-splines defined over exactly the same set of knots are identical. In essence, a B-spline is uniquely characterized by its knot sequence.

 \begin{algorithm*}[t!]
 \caption{The proposed EMI algorithm}
 	\label{alg:MPEC}
 	\begin{algorithmic}[1]
 		\Require Observations $(y_i^{(\text{off})},\bx_i^{(\text{off})}), i=1,2,\ldots,N^{(\text{off})}$, spline degree $d$, knots number $M$, moderate quantile level $\tau_0$, extreme quantile level $\tau_N$.
 		\Ensure Conditional quantile estimation $\widetilde Q_Y(\tau_N|\bx^{(\text{on})})$.
 		\\
 		\emph{Extrapolation.}
 		\While {$i=1,2,\cdots,N^{(\text{off})}$}
 		\State Solve MPEC problem~\eqref{eq:MPEC} with $\bx_i^{(\text{off})}$ and obtain $(\widetilde\alpha(\tau_0),\widetilde\bbeta(\tau_0),\widetilde\gamma(\bx_i^{(\text{off})}),\widetilde\sigma(\bx_i^{(\text{off})};\tau_0)))$. 
 		\EndWhile
 		\\
 		\emph{Interpolation.}
 		\State Calculate the knots $\left\{\widetilde{\xi}_{i,\gamma}\right\}_{i=1}^{M}$ for sigma via Eq~\eqref{eq:spline_gamma}.
 		\State Calculate the knots $\left\{\widetilde{\xi}_{i,\sigma}\right\}_{i=1}^{M}$ for sigma via Eq~\eqref{eq:spline_sigma}.
 		 \\
 		\emph{Prediction.}
 		 \State Calculate $\widetilde\gamma(\bx^{(\text{on})})$ by B-spline interpolation via Eq~\eqref{Bspline_gamma}.
 		 \State Calculate $\widetilde\sigma(\bx^{(\text{on})}; \tau_0)$ by B-spline interpolation via Eq~\eqref{Bspline_sigma}.
 		 \State Calculate $\widetilde Q_Y(\tau_N|\bx^{(\text{on})})$ via Eq.~\eqref{Bspline_plug-in}.
 	\end{algorithmic}
 \end{algorithm*}

B-splines provide a powerful method for approximating complex functions. In scenarios where the covariants are multi-dimensional, we estimate the covariate-dependent variables $\gamma(\bx)$ and the scale $\sigma(\bx,\tau_0)$ as
\begin{equation*}
	\begin{split}
		\gamma(\bx) &\approx \sum_{i=1}^p \sum_{j=1}^J B_{j, D, \boldsymbol{\xi}^{(i)}}\left(\bx(i)\right)t_{\gamma,j},\\
		\sigma(\bx,\tau_0) &\approx \sum_{i=1}^p \sum_{j=1}^J B_{j, D, \boldsymbol{\xi}^{(i)}}\left(\bx(i)\right)t_{\sigma,j}.
	\end{split}
\end{equation*}
Here, $\boldsymbol{\xi}^{(1)}, \boldsymbol{\xi}^{(2)}, \ldots, \boldsymbol{\xi}^{(p)}$ represent the $p$ sets of knots, with each set containing $M$ knots. It is assumed that $\bx(i)$ falls within the range defined by the boundary knots $\xi^{(i)}_1$ and $\xi^{(i)}_M$. The choice of $D$, representing the maximum degree of approximation, is crucial. While higher degrees tend to yield more accurate estimates, they also come with increased computational complexity. Here, we opt for $D=3$, a well-known choice commonly referred to as cubic spline interpolation. To determine the coefficients $t_{\gamma,j}, j = 1,\ldots,J$,  for $\mathbf{s} = [s_1, s_2, \ldots, s_N]^\top \in\mathbb{R}^N$, we introduce the B-spline matrix $\mathbf{B}_{d, \boldsymbol{\xi}}^{\mathbf{s}}\in\mathbb{R}^{N\times J}$ as
\begin{equation*}
\mathbf{B}_{d, \boldsymbol{\xi}}^{\mathbf{s}} = \left[\begin{array}{cccc}
B_{1, d, \boldsymbol{\xi}}(s_1) & B_{2, d, \boldsymbol{\xi}}(s_1) & \cdots & B_{J, d, \boldsymbol{\xi}}(s_1) \\ 
B_{1, d, \boldsymbol{\xi}}(s_2) & B_{2, d, \boldsymbol{\xi}}(s_2) & \cdots & B_{J, d, \boldsymbol{\xi}}(s_2) \\ 
\ldots & \ldots & \ldots & \ldots \\
B_{1, d, \boldsymbol{\xi}}(s_N) & B_{2, d, \boldsymbol{\xi}}(s_N) & \cdots & B_{J, d, \boldsymbol{\xi}}(s_N)\end{array}\right].
\end{equation*}
We further define the feature vector $\mathbf{v}_j$ as follows
\begin{equation*}
	\mathbf{v}_j = [\bx_1^{(\text{off})}(j), \bx_2^{(\text{off})}(j), \ldots, \bx_{N^{(\text{off})}}^{(\text{off})}(j)]^\top \in\mathbb{R}^{N^{(\text{off})}},
\end{equation*}
where $j  = 1,2,\ldots, p$ and $\bx_i^{^{(\text{off})}}(j)$ represents the $j$-th entry of observation $\bx_i^{^{(\text{off})}}$. For pairs $(\widetilde\gamma(\bx_i^{(\text{off})}),\bx_i^{^{(\text{off})}}), i=1,2,\ldots,N^{(\text{off})}$, we we can determine the coefficients $t_{\gamma,j}, j = 1,\ldots,J$ by solving the following matrix equation
\begin{equation}\label{eq:spline_gamma}
\left( \sum_{i=1}^p\mathbf{B}_{D, \boldsymbol{\xi}^{(i)}}^{\mathbf{v}_i} \right)
\left[\begin{array}{l}t_{\gamma,1} \\ t_{\gamma,2} \\ \ldots \\ t_{\gamma,J}\end{array}\right]
=\left[\begin{array}{c}\widetilde{\gamma}(\bx_1^{(\text{off})}) \\ \widetilde{\gamma}(\bx_2^{(\text{off})}) \\ \ldots \\ \widetilde{\gamma}(\bx_{N^{(\text{off})}}^{(\text{off})})\end{array}\right].
\end{equation}
with the boundary knots satisfy
\begin{equation*}
	{\xi}^{(i)}_1 = \underset{s}{\min}~\bx_s^{(\text{off})}(i),\quad{\xi}^{(i)}_M = \underset{s}{\max}~\bx_s^{(\text{off})}(i),
\end{equation*}
for $i = 1,\ldots, p$. The solution to this equation is denoted as $\widetilde{t}_{\gamma,j}, j = 1,\ldots,J$. Then for new observation $\bx^{{(\text{on})}}$, we can estimate the corresponding extreme value index by
\begin{equation*}
	\widetilde\gamma(\bx^{(\text{on})}) = \sum_{i=1}^p \sum_{j=1}^J B_{j, D, \boldsymbol{\xi}^{(i)}}\left(\bx^{(\text{on})}(i)\right)\widetilde{t}_{\gamma,j},
\end{equation*}
when $\boldsymbol{\xi}^{(i)}_1 \leq\bx^{(\text{on})}(i)\leq\boldsymbol{\xi}^{(i)}_M$. For cases where $\bx(i)$ falls beyond the range of boundary knots $\xi^{(i)}_1$ and $\xi^{(i)}_M$, the estimation of the extreme value index can be expressed as follows
\begin{equation*}
	\widetilde\gamma(\bx^{(\text{on})}) = \sum_{i=1}^p \sum_{j=1}^J B_{j, D, \boldsymbol{\xi}^{(i)}}\left(\widetilde{\bx}_\gamma(i)\right)\widetilde{t}_{\gamma,j},
\end{equation*}
where $\widetilde{\bx}_\gamma(i)$ is determined by minimizing the absolute difference as
\begin{equation*}
	\widetilde{\bx}_\gamma(i) = \arg\min_{s}\left|\bx^{(\text{on})}(i) - s\right|, s \in \{\bx_n^{(\text{off})}(i)\}_{n=1}^{N^{(\text{off})}}.
\end{equation*}
In summary, we estimate $\gamma(\bx)$ as
\begin{equation}\label{Bspline_gamma}\footnotesize
	\begin{split}
		&\widetilde\gamma(\bx^{(\text{on})}) \\
		=& \sum_{i=1}^p \sum_{j=1}^J\left\{\begin{array}{ll} B_{j, D, \boldsymbol{\xi}^{(i)}}\left(\bx^{(\text{on})}(i)\right)\widetilde{t}_{\gamma,j}, & \text { if }\boldsymbol{\xi}^{(i)}_1 \leq\bx^{(\text{on})}(i)\leq\boldsymbol{\xi}^{(i)}_M, \\ 
	B_{j, D, \boldsymbol{\xi}^{(i)}}\left(\widetilde{\bx}_\gamma(i)\right)\widetilde{t}_{\gamma,j}, & \text { otherwise }.\end{array}\right.
	\end{split}
\end{equation}

Similarly, the scale parameter $\sigma(\bx, \tau_0)$ can be estimated using B-spline interpolation as:
\begin{equation}\label{Bspline_sigma}\footnotesize
	\begin{split}
		&\widetilde\sigma(\bx^{(\text{on})}, \tau_0) \\
		=& \sum_{i=1}^p \sum_{j=1}^J\left\{\begin{array}{ll} B_{j, D, \boldsymbol{\xi}^{(i)}}\left(\bx^{(\text{on})}(i)\right)\widetilde{t}_{\sigma,j}, & \text { if }\boldsymbol{\xi}^{(i)}_1 \leq\bx^{(\text{on})}(i)\leq\boldsymbol{\xi}^{(i)}_M, \\ 
	B_{j, D, \boldsymbol{\xi}^{(i)}}\left(\widetilde{\bx}_\sigma(i)\right)\widetilde{t}_{\sigma,j}, & \text { otherwise }.\end{array}\right.
	\end{split}
\end{equation}
In this expression, the coefficients $\widetilde{t}_{\sigma,j}$ for $j=1,2,\ldots,J$ are determined by solving the matrix equation given by:
\begin{equation}\label{eq:spline_sigma}
\left( \sum_{i=1}^p\mathbf{B}_{D, \boldsymbol{\xi}^{(i)}}^{\mathbf{v}_i} \right)
\left[\begin{array}{l}t_{\sigma,1} \\ t_{\sigma,2} \\ \ldots \\ t_{\sigma,J}\end{array}\right]
=\left[\begin{array}{c}\widetilde{\sigma}(\bx_1^{(\text{off})},\tau_0) \\ \widetilde{\sigma}(\bx_2^{(\text{off})},\tau_0) \\ \ldots \\ \widetilde{\sigma}(\bx_{N^{(\text{off})}}^{(\text{off})},\tau_0)\end{array}\right],
\end{equation}
and $\widehat{\bx}_\sigma(i)$ satisfying 
\begin{equation*}
	\widetilde{\bx}_\sigma(i) = \arg\min_{s}\left|\bx^{(\text{on})}(i) - s\right|, s \in \{\bx_n^{(\text{off})}(i)\}_{n=1}^{N^{(\text{off})}}.
\end{equation*}
In this way, the estimation of extreme conditional quantile for observation $\bx^{(\text{on})}$ is calculated as
\begin{equation}\label{Bspline_plug-in}\footnotesize
	\begin{split}
		& \widetilde Q_Y(\tau_0|\bx^{(\text{on})})=\widetilde\alpha(\tau_0)+\widetilde\bbeta^T(\tau_0)\bx^{(\text{on})}, \\
		& \widetilde Q_Y(\tau_N|\bx^{(\text{on})})\\
		=&\left\{
		\begin{matrix}\widetilde Q_Y(\tau_N|\bx^{(\text{on})})+\frac{\widetilde\sigma(\bx^{(\text{on})},\tau_0)}{\widetilde\gamma(\bx^{(\text{on})})}\left(\tau_{0,N}^{\widetilde\gamma(\bx^{(\text{on})})}-1\right), & \widetilde\gamma(\bx^{(\text{on})})\ne 0, \\ \widetilde Q_Y(\tau_0|\bx^{(\text{on})})+\widetilde\sigma(\bx^{(\text{on})},\tau_0)\log\tau_{0,N}, & \widetilde\gamma(\bx^{(\text{on})})=0.
		\end{matrix}\right.
	\end{split}
\end{equation}
We summarize our method in Algorithm~\ref{alg:MPEC}. 

It is important to emphasize that B-spline interpolation, while highly effective for approximating scale and shape parameters, does introduce a degree of bias into the estimation process. This trade-off is essential as it significantly reduces computational time, making it a valuable tool for online observations. For this specific context, there is no need to engage in solving the censored maximum likelihood problem required for GPD distribution fitting when dealing with online data. Instead, we capitalize on historical offline data to perform the approximation using B-spline interpolation. It is noteworthy that the solutions derived from equations like~\eqref{eq:spline_gamma} and~\eqref{eq:spline_sigma} exclusively rely on the historical dataset. Consequently, we can obtain these solutions well in advance of the arrival of new online data. This proactive approach implies that once the online data is observed, we can directly substitute the results into~\eqref{Bspline_plug-in} to efficiently obtain the parameter estimations, thus drastically reducing the computational burden associated with online data processing.

\section{Experiments}
 \begin{figure*}[ht!]
	\centering
	\subfigure[$N^{(\text{on})}=1000,p=10$]{
		\begin{minipage}[t]{0.23\linewidth}
				\includegraphics[scale = 0.2]{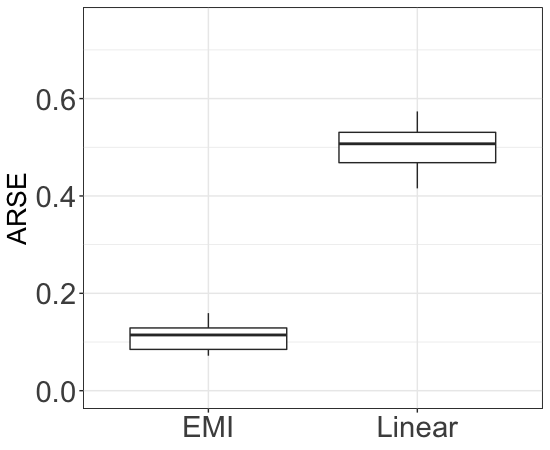}
		\end{minipage}%
	}
	\subfigure[$N^{(\text{on})}=2000,p=10$]{
		\begin{minipage}[t]{0.23\linewidth}
				\includegraphics[scale = 0.2]{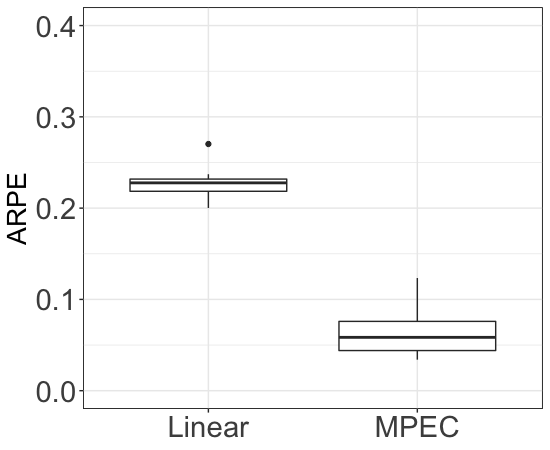}
		\end{minipage}%
	}
		\subfigure[$N^{(\text{on})}=2000,p=20$]{
		\begin{minipage}[t]{0.23\linewidth}
				\includegraphics[scale = 0.2]{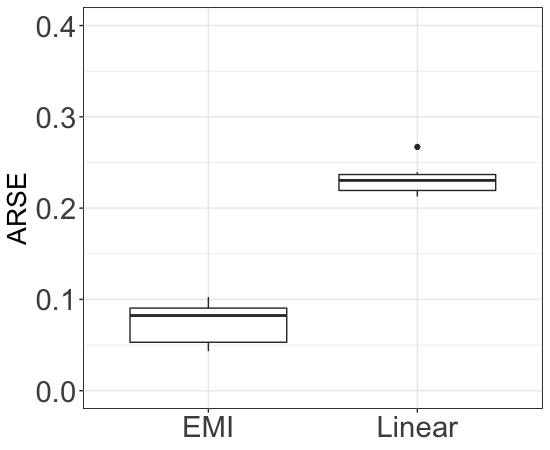}
		\end{minipage}%
	}
	\subfigure[$N^{(\text{on})}=2000,p=30$]{
		\begin{minipage}[t]{0.23\linewidth}
				\includegraphics[scale = 0.2]{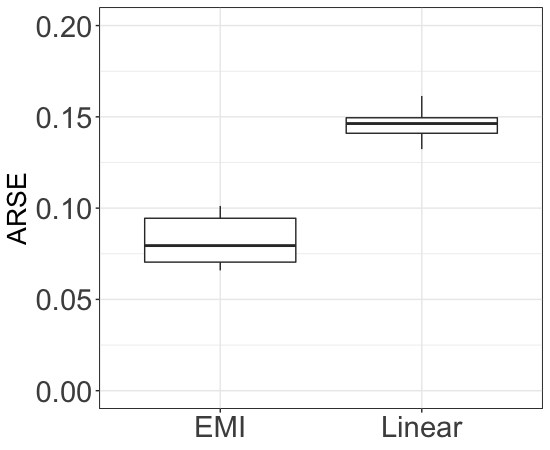}
		\end{minipage}%
	}
	\vskip 1em

	\subfigure[$N^{(\text{on})}=1000,p=10$]{
		\begin{minipage}[t]{0.23\linewidth}
				\includegraphics[scale = 0.2]{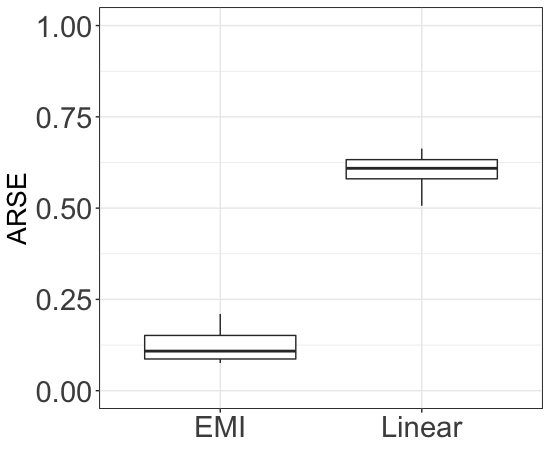}
		\end{minipage}%
	}
	\subfigure[$N^{(\text{on})}=2000,p=10$]{
		\begin{minipage}[t]{0.23\linewidth}
				\includegraphics[scale = 0.2]{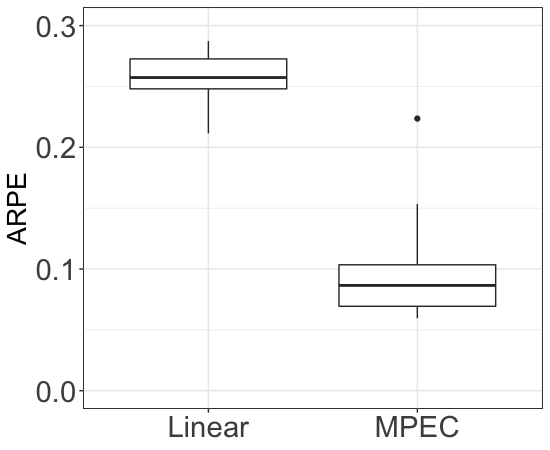}
		\end{minipage}%
	}
		\subfigure[$N^{(\text{on})}=2000,p=20$]{
		\begin{minipage}[t]{0.23\linewidth}
				\includegraphics[scale = 0.2]{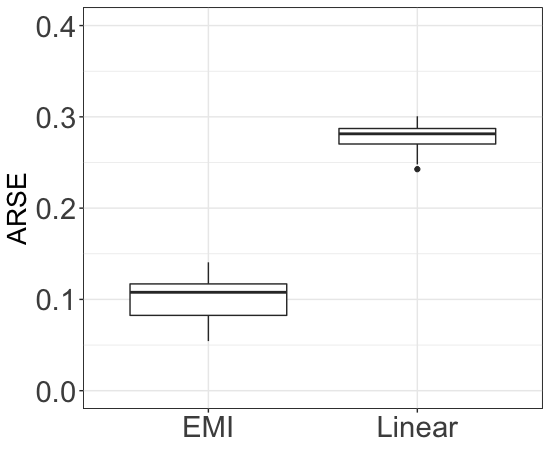}
		\end{minipage}%
	}
	\subfigure[$N^{(\text{on})}=2000,p=30$]{
		\begin{minipage}[t]{0.23\linewidth}
				\includegraphics[scale = 0.2]{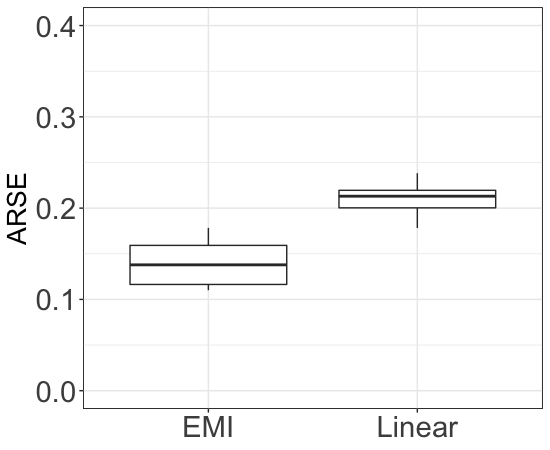}
		\end{minipage}%
	}
	\vskip 1em

	\subfigure[$N^{(\text{on})}=1000,p=10$]{
		\begin{minipage}[t]{0.23\linewidth}
				\includegraphics[scale = 0.2]{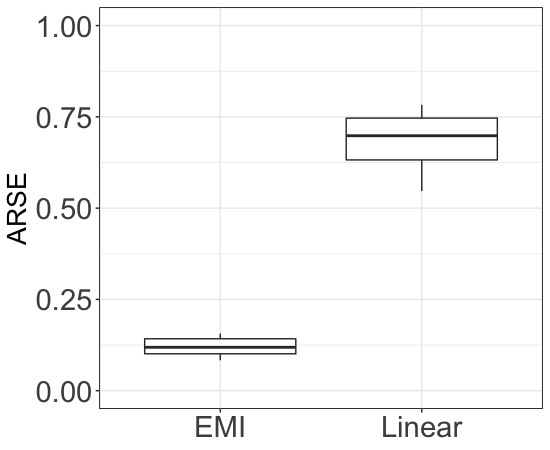}
		\end{minipage}%
	}
	\subfigure[$N^{(\text{on})}=2000,p=10$]{
		\begin{minipage}[t]{0.23\linewidth}
				\includegraphics[scale = 0.2]{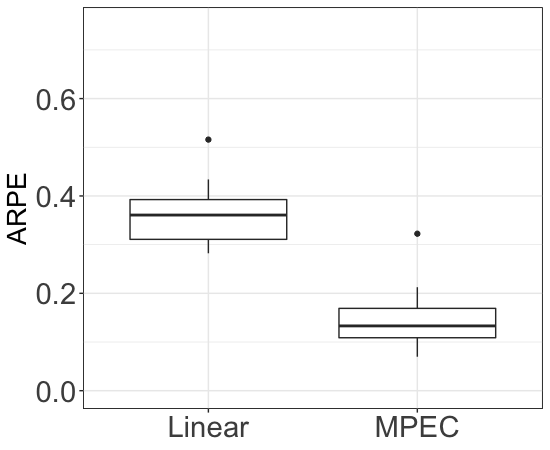}
		\end{minipage}%
	}
		\subfigure[$N^{(\text{on})}=2000,p=20$]{
		\begin{minipage}[t]{0.23\linewidth}
				\includegraphics[scale = 0.2]{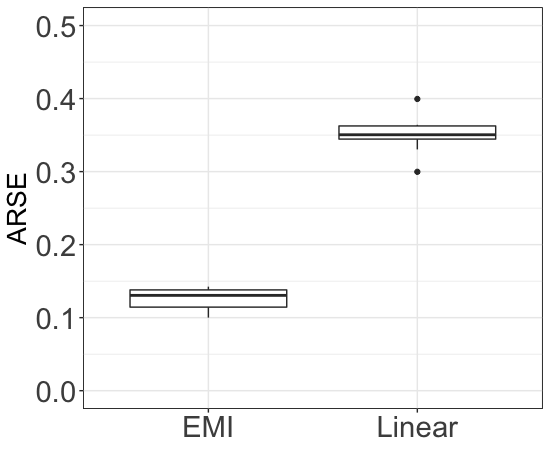}
		\end{minipage}%
	}
	\subfigure[$N^{(\text{on})}=2000,p=30$]{
		\begin{minipage}[t]{0.23\linewidth}
				\includegraphics[scale = 0.2]{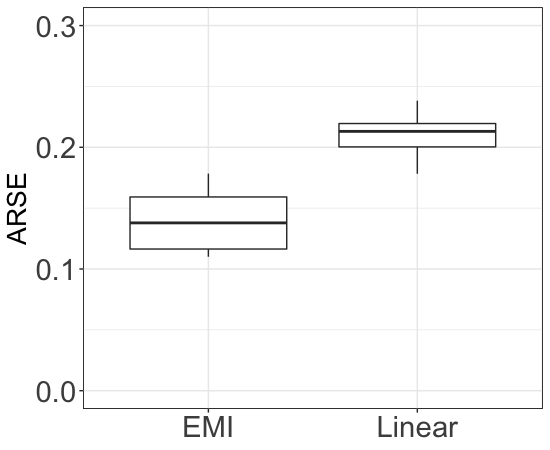}
		\end{minipage}%
	}
	\caption{The boxplot of ARSE for different methods with different extreme quantile levels. The top, middle, and below rows are the results of $\tau_N=0.99$, $\tau_N=0.995$, and $\tau_N=0.999$ respectively. The noise variable $\epsilon_i$ is drawn from the t-distribution.}
	\label{fig:1}
\end{figure*}

 \begin{figure*}[h]
	\centering
	\subfigure[$N^{(\text{on})}=1000,p=10$]{
		\begin{minipage}[t]{0.23\linewidth}
				\includegraphics[scale = 0.2]{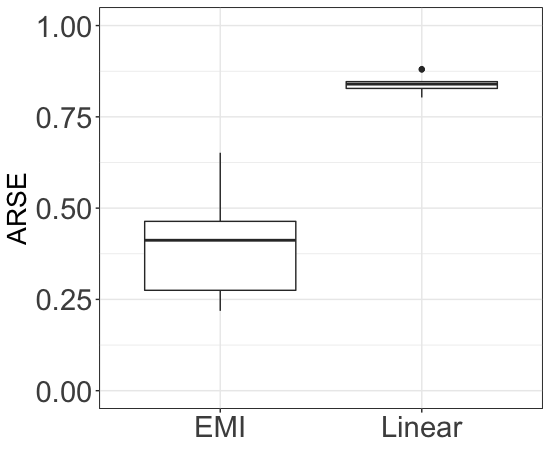}
		\end{minipage}%
	}
	\subfigure[$N^{(\text{on})}=2000,p=10$]{
		\begin{minipage}[t]{0.23\linewidth}
				\includegraphics[scale = 0.2]{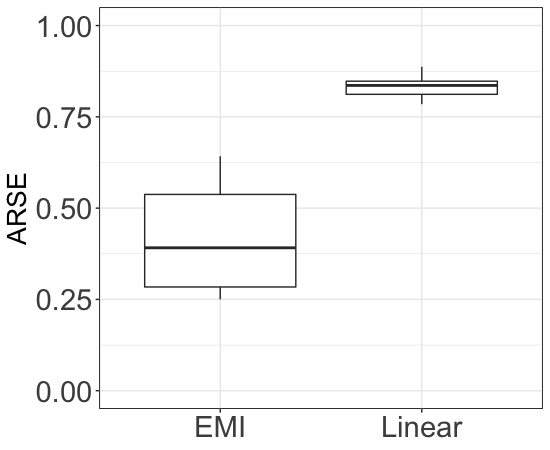}
		\end{minipage}%
	}
		\subfigure[$N^{(\text{on})}=2000,p=20$]{
		\begin{minipage}[t]{0.23\linewidth}
				\includegraphics[scale = 0.2]{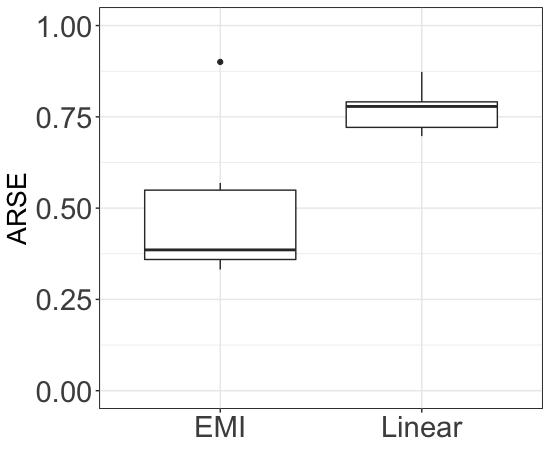}
		\end{minipage}%
	}
	\subfigure[$N^{(\text{on})}=2000,p=30$]{
		\begin{minipage}[t]{0.23\linewidth}
				\includegraphics[scale = 0.2]{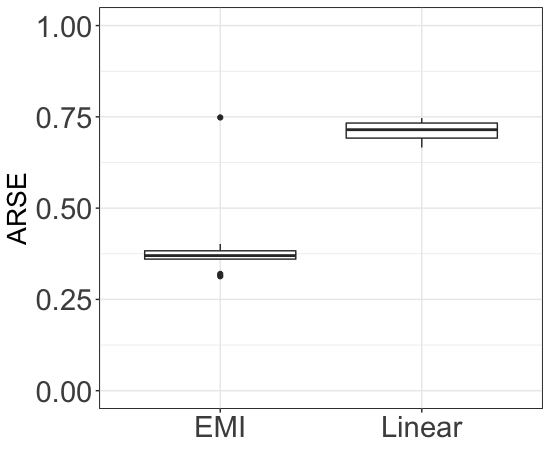}
		\end{minipage}%
	}
	\vskip 1em

	\subfigure[$N^{(\text{on})}=1000,p=10$]{
		\begin{minipage}[t]{0.23\linewidth}
				\includegraphics[scale = 0.2]{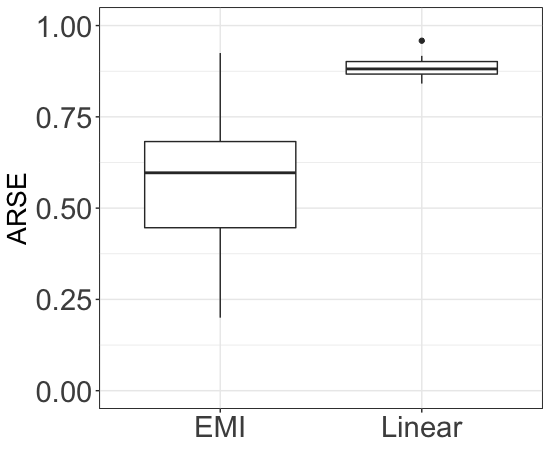}
		\end{minipage}%
	}
	\subfigure[$N^{(\text{on})}=2000,p=10$]{
		\begin{minipage}[t]{0.23\linewidth}
				\includegraphics[scale = 0.2]{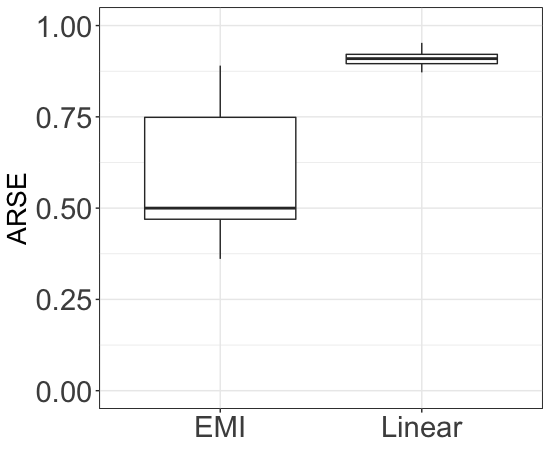}
		\end{minipage}%
	}
		\subfigure[$N^{(\text{on})}=2000,p=20$]{
		\begin{minipage}[t]{0.23\linewidth}
				\includegraphics[scale = 0.2]{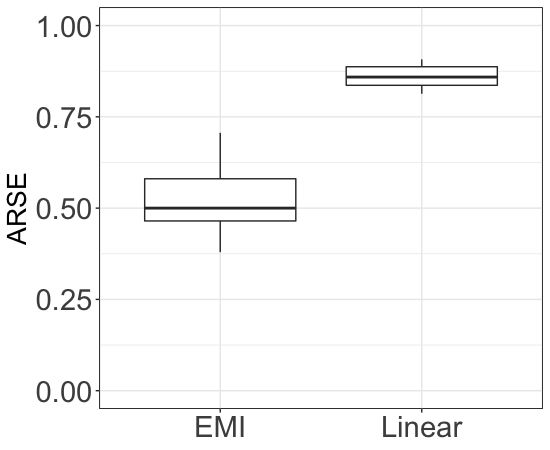}
		\end{minipage}%
	}
	\subfigure[$N^{(\text{on})}=2000,p=30$]{
		\begin{minipage}[t]{0.23\linewidth}
				\includegraphics[scale = 0.2]{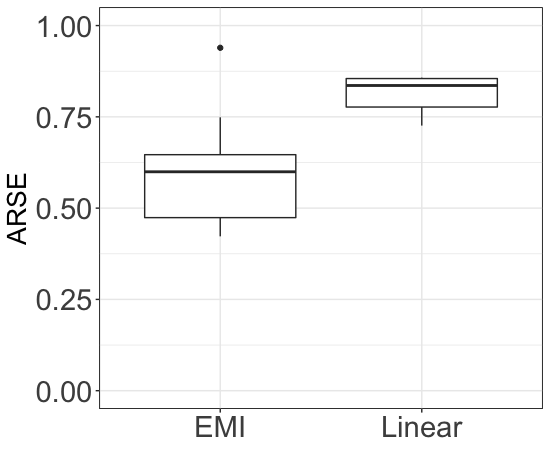}
		\end{minipage}%
	}
	\vskip 1em

	\subfigure[$N^{(\text{on})}=1000,p=10$]{
		\begin{minipage}[t]{0.23\linewidth}
				\includegraphics[scale = 0.2]{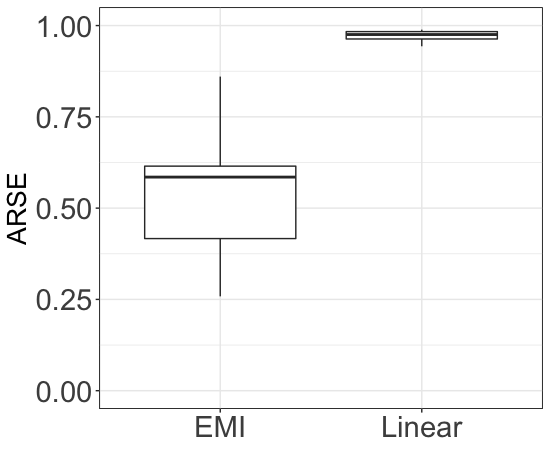}
		\end{minipage}%
	}
	\subfigure[$N^{(\text{on})}=2000,p=10$]{
		\begin{minipage}[t]{0.23\linewidth}
				\includegraphics[scale = 0.2]{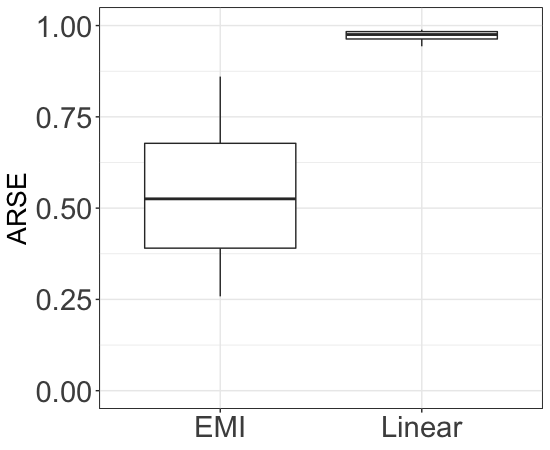}
		\end{minipage}%
	}
		\subfigure[$N^{(\text{on})}=2000,p=20$]{
		\begin{minipage}[t]{0.23\linewidth}
				\includegraphics[scale = 0.2]{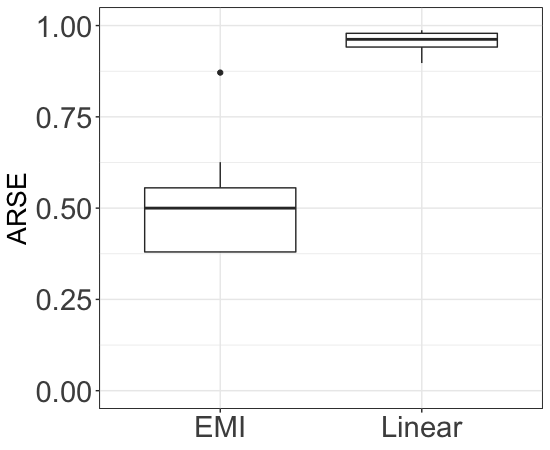}
		\end{minipage}%
	}
	\subfigure[$N^{(\text{on})}=2000,p=30$]{
		\begin{minipage}[t]{0.23\linewidth}
				\includegraphics[scale = 0.2]{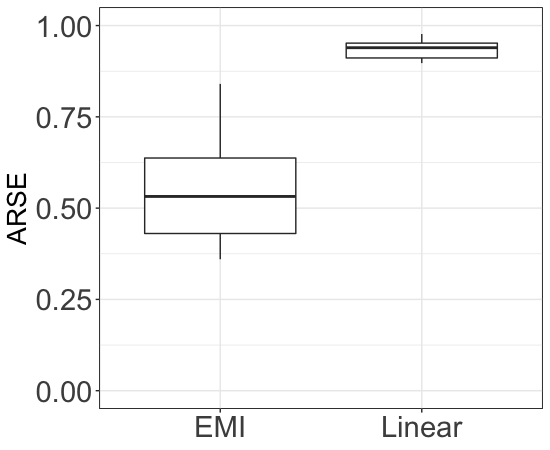}
		\end{minipage}%
	}
	\caption{The boxplot of the ARSE for different methods with different extreme quantile levels. The top, middle, and below rows are the results of $\tau_N=0.99$, $\tau_N=0.995$, and $\tau_N=0.999$ respectively. The noise variable $\epsilon_i$ is drawn from the GPD.}
	\label{fig:2}
\end{figure*}

\subsection{Setup}
In this section, we conduct some simulations to investigate the performance of our proposed EMI. We generate offline data $(y_i^{(\text{off})},\bx_i^{(\text{off})}),i=1,2,\ldots,N^{(\text{off})}$ and online data $(y_i^{(\text{on})},\bx_i^{(\text{on})}),i=1,2,\ldots,N^{(\text{on})}$ in the same way. Specifically, we generate the covariates $\bx_i \in\mathbb{R}^p $ independently from the standard uniform distribution. The response $y_i$ is generated by
	\begin{equation*}
		y_i = \alpha + \bbeta^\top \bx_i + \epsilon_i,
	\end{equation*}
where we consider two models of the univariate variable $\epsilon_i$:
\begin{itemize}
	\item Model 1: $\epsilon_i$ is the absolute value of a random variable drawn from the t-distribution with the zero location and degree $\psi_i$ satisfying
	\begin{equation*}
		\epsilon_i \sim \operatorname{T}(\psi), \quad \psi_i = \operatorname{exp}\left(\boldsymbol{\eta}_1^\top \bx_i\right).
	\end{equation*}
	
	\item Model 2: $\epsilon_i$ is the absolute value of a random variable drawn from the GPD with the location equal to $1$, the scale parameter equal to $0.25$, and the shape parameter $\phi_i$ satisfying
	\begin{equation*}
		\epsilon_i \sim \operatorname{GPD}(1,0.25,\phi_i), \quad \phi_i = 1/(\boldsymbol{\eta}_2^\top \bx_i).
	\end{equation*}
\end{itemize}
Note that both models are heavy-tailed, where the degrees of t-distribution and the shape parameters of GPD all depend on the covariates. We set $\boldsymbol{\eta}_1 = [0.2, \ldots, 0.2]^\top\in\mathbb{R}^p$ and $\boldsymbol{\eta}_2 = [0.6, \ldots, 0.6]^\top\in\mathbb{R}^p$. Without loss of generality, we set $\alpha = 0.5$ and $\bbeta = [0.5, \ldots, 0.5]^\top\in\mathbb{R}^p$. We aim to estimate the conditional quantile function $Q_Y(\tau_N|\bx)$ corresponding to extreme probability levels $\tau_N \in \{0.99,0.995,0.999\}$ and $N^{(\text{off})} = 1000$. The intermediate quantile level $\tau_0$ is set to $0.8$.

To assess the performance, we conduct $100$ replications and calculate the average relative prediction square error (ARSE) on the online data $(y_i^{(\text{on})},\bx_i^{(\text{on})}),i=1,2,\ldots,N^{(\text{on})}$. For the $j$-th replication, the ARSE$\left(\tau_N, j\right)$ is defined as
	\begin{equation*}
		\text{ARSE}\left(\tau_N, j\right) = \frac{1}{N^{(\text{on})}}\sum_{i=1}^{N^{(\text{on})}}\left(\frac{\widetilde Q_Y(\tau_N|\bx_i^{(\text{on})})}{Q_Y(\tau_N|\bx_i^{(\text{on})})} -1\right)^2.
	\end{equation*}
	Here $\widetilde Q_Y(\tau_N|\bx_i^{(\text{on})})$ is the estimator calculated according to Eq.~\eqref{Bspline_plug-in} and $Q_Y(\tau_N|\bx_i^{(\text{on})})$ is the true conditional quantile.  For comparison, we apply the traditional method (denoted as \emph{Linear}) which solves the following problem
	\begin{equation}\label{linear}
		\min _{\alpha,\bbeta} \sum_{i=1}^{N^{(\text{on})}}\rho_{\tau}(y_i^{(\text{on})}-\alpha-\bbeta^T\bx_i^{(\text{on})}),
	\end{equation}
	with $\tau = \tau_N$. \emph{Linear} estimates the conditional quantile as the linear function of $\bx_i^{(\text{on})}$:
 	\begin{equation*}\small
 		\widetilde Q_Y'(\tau_N|\bx_i^{(\text{on})})=\widetilde{\alpha}'(\tau_0)+\widetilde{\bbeta}'^\top(\tau_0)\bx_i^{(\text{on})},~ i = 1,\ldots, N^{(\text{on})},
 	\end{equation*}
 	where $\widetilde{\alpha}'(\tau_N)$ and $\widetilde{\bbeta}'(\tau_N)$ denote the solutions to~\eqref{linear}.

\begin{figure*}[t!]
	\centering
	\subfigure[Offline data size.]{
		\begin{minipage}[t]{0.45\linewidth}
				\includegraphics[scale = 0.36]{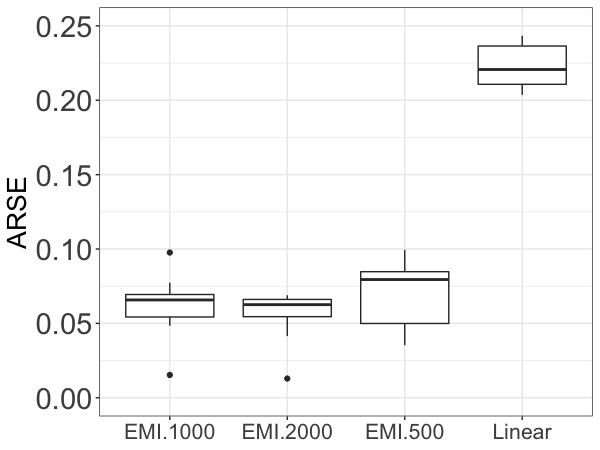}
		\end{minipage}%
		\label{fig:3a}
	}
	\subfigure[Choice of threshold $\tau_0$.]{
		\begin{minipage}[t]{0.45\linewidth}
				\includegraphics[scale = 0.36]{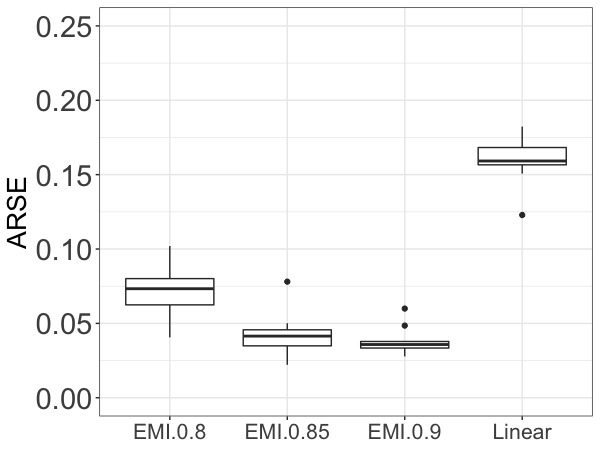}
		\end{minipage}%
		\label{fig:3b}
	}
	\caption{The boxplot of the ARSE with $N^{(\text{on})}=1000$, $p=30$. The noise variable $\epsilon_i$ is drawn from the t-distribution. (a) The ARSE under different offline data sizes with $\tau_N = 0.999$. (b) The ARSE under different choices of threshold $\tau_0$ with $\tau_N = 0.99$.}
\end{figure*}

\subsection{Estimation Results}	

The comparison of our proposed EMI to the \emph{Linear} approach is presented in Fig.~\ref{fig:1} and Fig.~\ref{fig:2}. For each chosen extreme quantile $\tau_N$, we generated data with various combinations of online data size $N^{(\text{on})}$ and feature dimension $p$. We observe that when the quantile level $\tau_N$ is close to $1$, the ARSE of both methods grows. This behavior is expected since, in such cases, observations above the $\tau_N$ quantile become scarce. However, our EMI clearly outperforms the \emph{Linear} approach with a significantly lower ARSE. This superior performance can be attributed to two key factors. Firstly, EMI leverages the GPD to approximate the tail distribution of the data, enabling extrapolation beyond the observation range. Secondly, when sufficient offline historical data is available, B-spline interpolation can effectively capture the relationship between $\bx$ and covariate-dependent parameters, resulting in better interpolation outcomes.

To further analyze the impact of the offline data size on estimation results, we conduct experiments with different offline data sizes ranging from $500$ to $2000$, as shown in Fig.~\ref{fig:3a}. The results reveal that, as the offline data size increases, the ARSE of our EMI gradually decreases while exhibiting reduced variance. This trend is due to the fact that larger data size can improve the performance of B-spline interpolation, particularly in cases where $\bx^{(\text{on})}_i$ falls beyond the observation range. Additionally, we examine the influence of different thresholds $\tau_0$ on the estimation. Fig.~\ref{fig:3b} shows that across various threshold choices, our EMI consistently exhibits superior estimation results compared to the \emph{Linear} approach.

\begin{figure*}[t!]
	\centering
	\subfigure[$\tau_N=0.99$]{
		\begin{minipage}[t]{0.32\linewidth}
				\includegraphics[scale = 0.26]{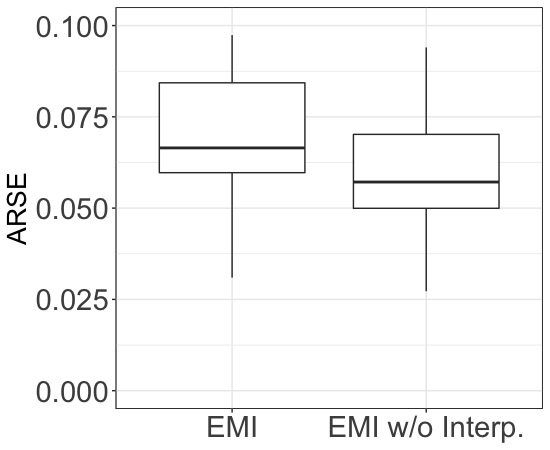}
		\end{minipage}%
	}
		\subfigure[$\tau_N=0.995$]{
		\begin{minipage}[t]{0.32\linewidth}
				\includegraphics[scale = 0.26]{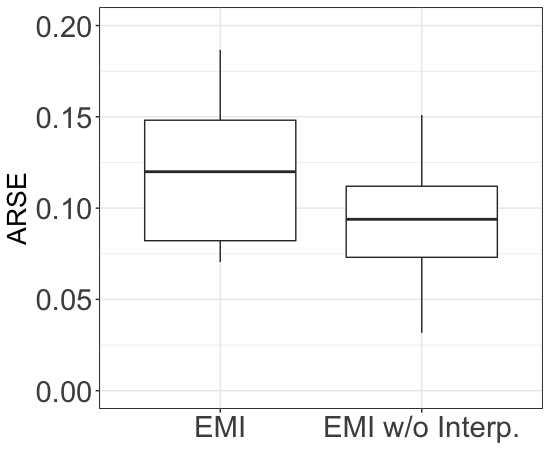}
		\end{minipage}%
	}
	\subfigure[$\tau_N=0.999$]{
		\begin{minipage}[t]{0.32\linewidth}
				\includegraphics[scale = 0.26]{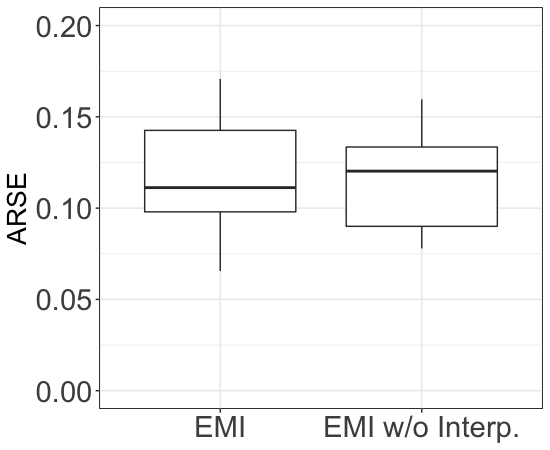}
		\end{minipage}%
	}
	\caption{The boxplot of the ARSE for different methods with different extreme quantile levels. The noise variable $\epsilon_i$ is drawn from the t-distribution. We set $N^{(\text{on})}=2000$ and $p=10$.}
		\label{fig:4}
\end{figure*}

\begin{figure*}[t!]
	\centering
	\subfigure[$\tau'=0.99$]{
		\begin{minipage}[t]{0.32\linewidth}
				\includegraphics[scale = 0.26]{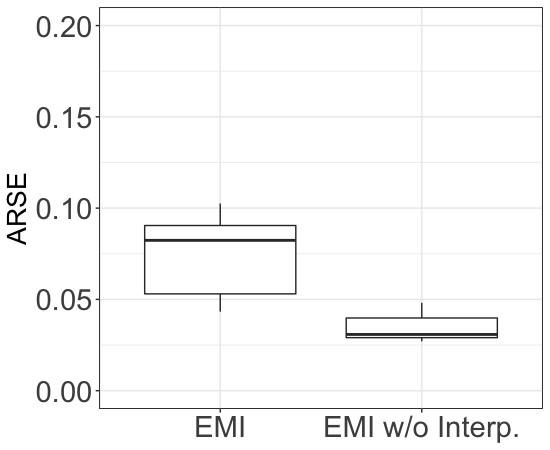}
		\end{minipage}%
	}
		\subfigure[$\tau'=0.995$]{
		\begin{minipage}[t]{0.32\linewidth}
				\includegraphics[scale = 0.26]{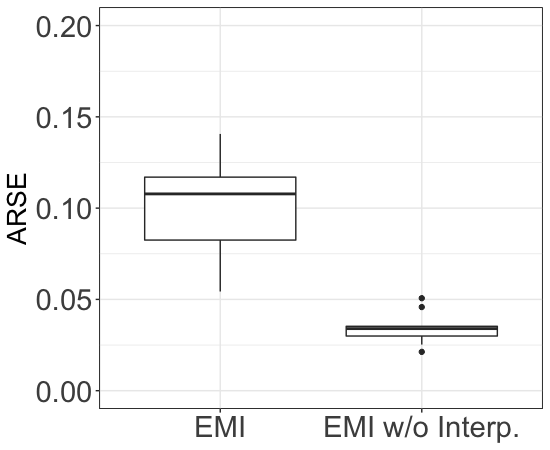}
		\end{minipage}%
	}
	\subfigure[$\tau'=0.999$]{
		\begin{minipage}[t]{0.32\linewidth}
				\includegraphics[scale = 0.26]{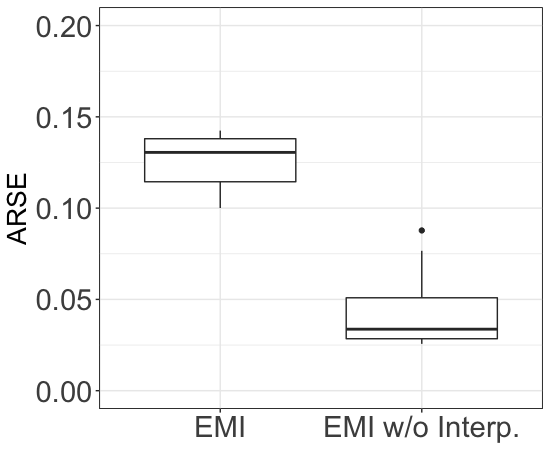}
		\end{minipage}%
	}
	\caption{The boxplot of the ARSE for different methods with different extreme quantile levels. The noise variable $\epsilon_i$ is drawn from the t-distribution. We set $N^{(\text{on})}=2000$ and $p=20$.}
		\label{fig:5}
\end{figure*}

\begin{figure*}[t!]
	\centering
	\subfigure[$\tau'=0.99$]{
		\begin{minipage}[t]{0.32\linewidth}
				\includegraphics[scale = 0.26]{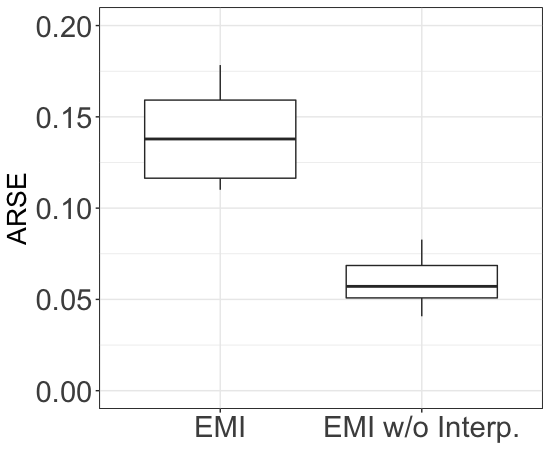}
		\end{minipage}%
	}
		\subfigure[$\tau'=0.995$]{
		\begin{minipage}[t]{0.32\linewidth}
				\includegraphics[scale = 0.26]{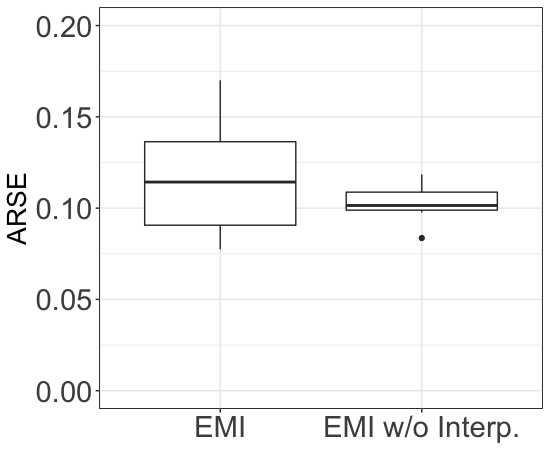}
		\end{minipage}%
	}
	\subfigure[$\tau'=0.999$]{
		\begin{minipage}[t]{0.32\linewidth}
				\includegraphics[scale = 0.26]{Fig4_5_6_Bias/n2000m30tau0.995.png}
		\end{minipage}%
	}
	\caption{The boxplot of the ARSE for different methods with different extreme quantile levels. The noise variable $\epsilon_i$ is drawn from the t-distribution. We set $N^{(\text{on})}=2000$ and $p=30$.}
		\label{fig:6}
\end{figure*} 

\begin{figure*}[t!]
	\centering
	\subfigure[$N^{(\text{on})}=2000,p=10$]{
		\begin{minipage}[t]{0.45\linewidth}
				\includegraphics[scale = 0.31]{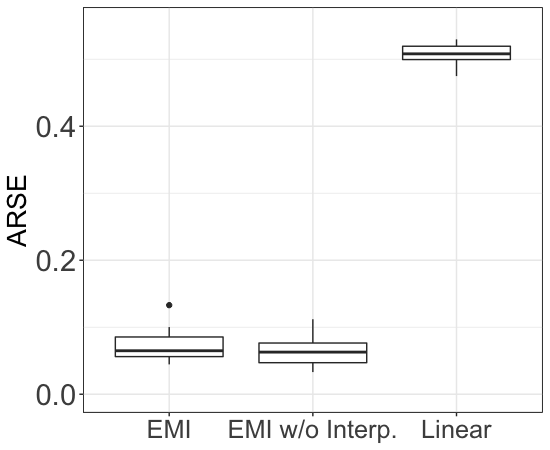}
		\end{minipage}%
	}
		\subfigure[$N^{(\text{on})}=2000,p=20$]{
		\begin{minipage}[t]{0.45\linewidth}
				\includegraphics[scale = 0.31]{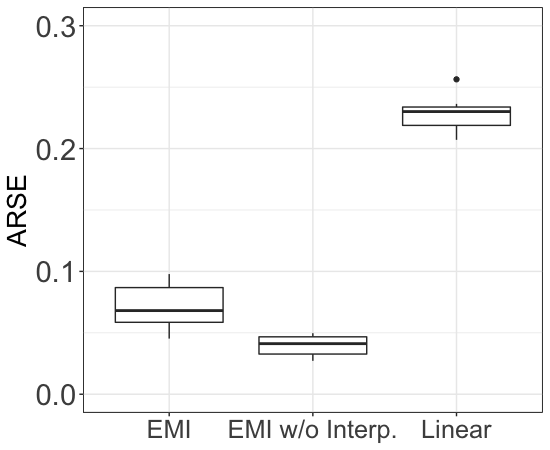}
		\end{minipage}%
	}
	
	\subfigure[$N^{(\text{on})}=2000,p=30$]{
		\begin{minipage}[t]{0.45\linewidth}
				\includegraphics[scale = 0.31]{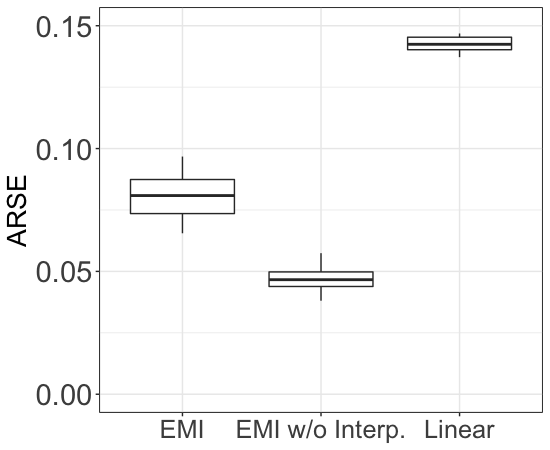}
		\end{minipage}%
	}
		\subfigure[$N^{(\text{on})}=2000,p=10$]{
		\begin{minipage}[t]{0.45\linewidth}
				\includegraphics[scale = 0.31]{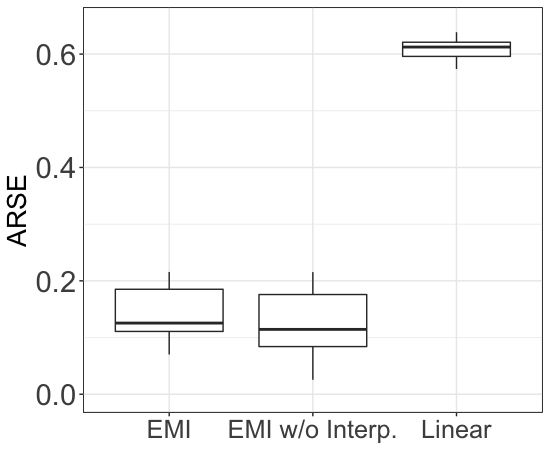}
		\end{minipage}%
	}
	\caption{The boxplot of the ARSE for different methods with different extreme quantile levels. We set $N^{(\text{off})}=1000$. (a)-(c) The ARSE under different covariate dimensions with $\tau_N = 0.99$. (d) The ARSE under different covariate dimensions with $\tau_N = 0.995$.}
		\label{fig:7a}
\end{figure*}

\begin{figure*}[t!]
	\centering
		\subfigure[$N^{(\text{on})}=2000,p=20$]{
		\begin{minipage}[t]{0.45\linewidth}
				\includegraphics[scale = 0.31]{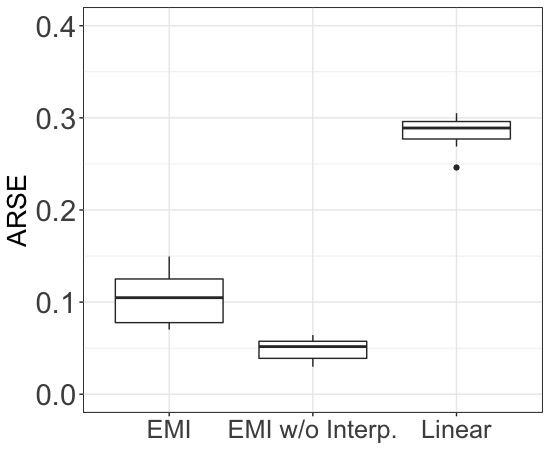}
		\end{minipage}%
	}
		\subfigure[$N^{(\text{on})}=2000,p=30$]{
		\begin{minipage}[t]{0.45\linewidth}
				\includegraphics[scale = 0.31]{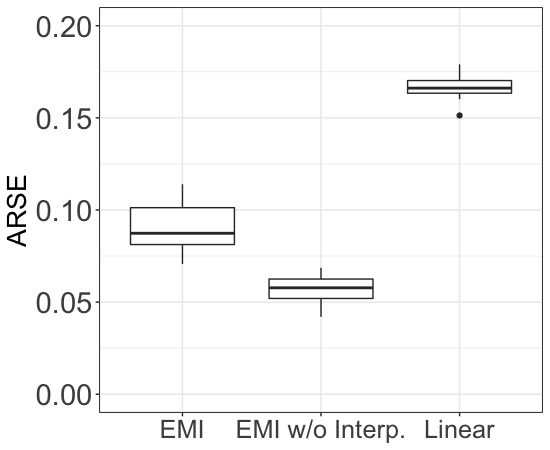}
		\end{minipage}%
	}
	
	\caption{The boxplot of the ARSE for different methods with $\tau_N = 0.995$. We set $N^{(\text{off})}=1000$.}
		\label{fig:7b}
\end{figure*}

\begin{figure*}[t!]
	\centering
	\subfigure[$N^{(\text{on})}=2000,p=10$]{
		\begin{minipage}[t]{0.45\linewidth}
				\includegraphics[scale = 0.31]{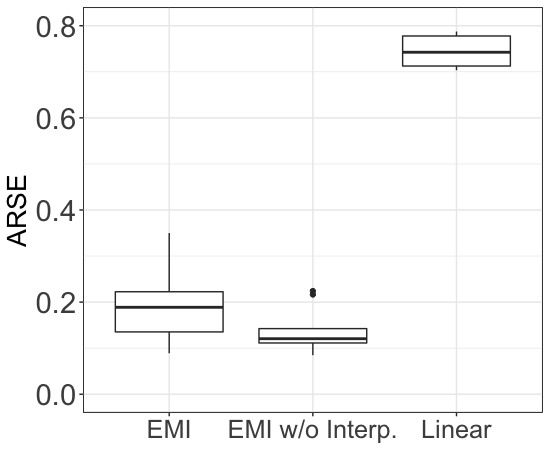}
		\end{minipage}%
	}
		\subfigure[$N^{(\text{on})}=2000,p=20$]{
		\begin{minipage}[t]{0.45\linewidth}
				\includegraphics[scale = 0.31]{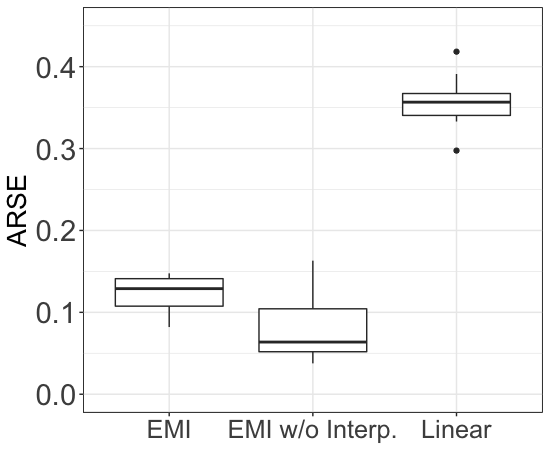}
		\end{minipage}%
	}
	
	\subfigure[$N^{(\text{on})}=2000,p=30$]{
		\begin{minipage}[t]{0.45\linewidth}
				\includegraphics[scale = 0.31]{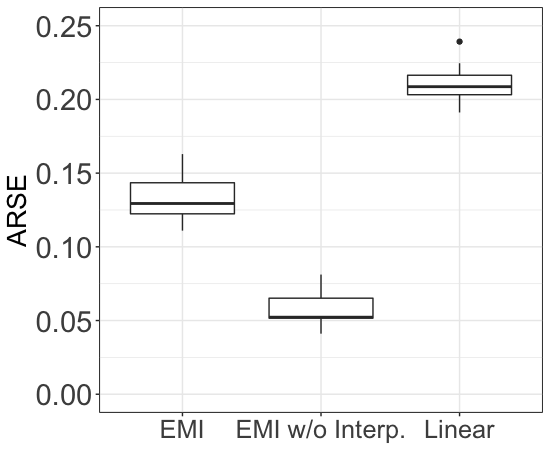}
		\end{minipage}%
	}
	\caption{The boxplot of the ARSE for different methods with different covariate dimensions. We set $N^{(\text{off})}=1000$ and $\tau_N = 0.995$.}
		\label{fig:8}
\end{figure*}

\subsection{Bias Analysis}

We then study the bias introduced by the B-spline interpolation. For comparison, given each online observation $\bx^{(\text{on})}_i$, we do not apply B-spline interpolation but fit the tail distribution into a GPD and estimate the shape and scale parameters, respectively. To be specific, we directly solve the following bilevel programming
\begin{equation}\label{opt:bilevel2}
	\begin{split}
		\max_{\gamma,\sigma} \quad &L^{(\text{MLE})}_{N}(\gamma,\sigma|Q_Y(\tau_0|\bx^{(\text{on})}_i))\\
		s.t.\quad & \min_{\alpha,\bbeta} L^{(\text{QR})}_{N\tau_0}(\alpha,\bbeta).
	\end{split}
\end{equation}
Denote the solutions to problem~\eqref{opt:bilevel2} as $(\bar{\alpha}(\tau_0), \bar{\bbeta}(\tau_0), \bar{\gamma}(\bx_i^{(\text{on})}),\bar{\sigma}(\bx_i^{(\text{on})};\tau_0)) $. 
Then the conditional quantile is estimated as
\begin{equation}\footnotesize
\begin{split}
		& \bar{Q}_Y(\tau_0|\bx^{(\text{on})})=\bar{\alpha}(\tau_0)+\bar{\bbeta}^T(\tau_0)\bx^{(\text{on})}, \\
		& \bar{Q}_Y(\tau_N|\bx^{(\text{on})})\\
		=&\left\{
		\begin{matrix}\bar{Q}_Y(\tau_N|\bx^{(\text{on})})+\frac{\bar{\sigma}(\bx^{(\text{on})},\tau_0)}{\bar{\gamma}(\bx^{(\text{on})})}\left(\tau_{0,N}^{\bar{\gamma}(\bx^{(\text{on})})}-1\right), & \bar{\gamma}(\bx^{(\text{on})})\ne 0, \\ \bar{Q}_Y(\tau_0|\bx^{(\text{on})})+\bar{\sigma}(\bx^{(\text{on})},\tau_0)\log\tau_{0,N}, & \bar{\gamma}(\bx^{(\text{on})})=0.
		\end{matrix}\right.
	\end{split}
\end{equation}
We denote the above method as \emph{EMI w/o. Interp.}, denoting EMI without B-spline interpolation.  We conducte performance tests under various extreme quantile levels $\tau_N$, and the results are summarized in Fig.\ref{fig:4}, Fig.\ref{fig:5}, and Fig.~\ref{fig:6}. Notably, we observe that EMI performs slightly less effectively when employing B-spline interpolation, which coincides with our expectations. However, it is important to emphasize that B-spline interpolation allows us to bypass the complex computations involved in fitting the GPD, which leads to more practical applications.

\begin{figure*}[t!]
	\centering
	\subfigure[$\tau_N = 0.99$]{
		\includegraphics[scale = 0.23]{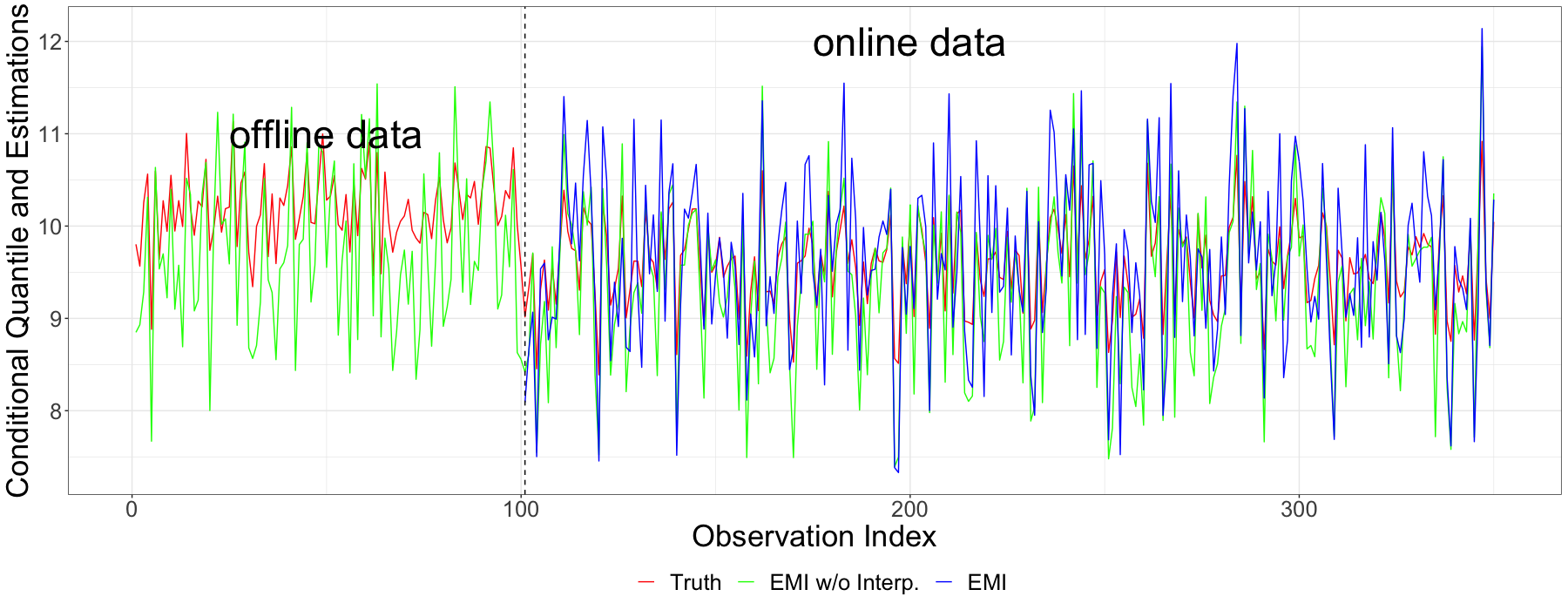}
		\centering
	}
	
	\subfigure[$\tau_N = 0.995$]{
		\includegraphics[scale = 0.23]{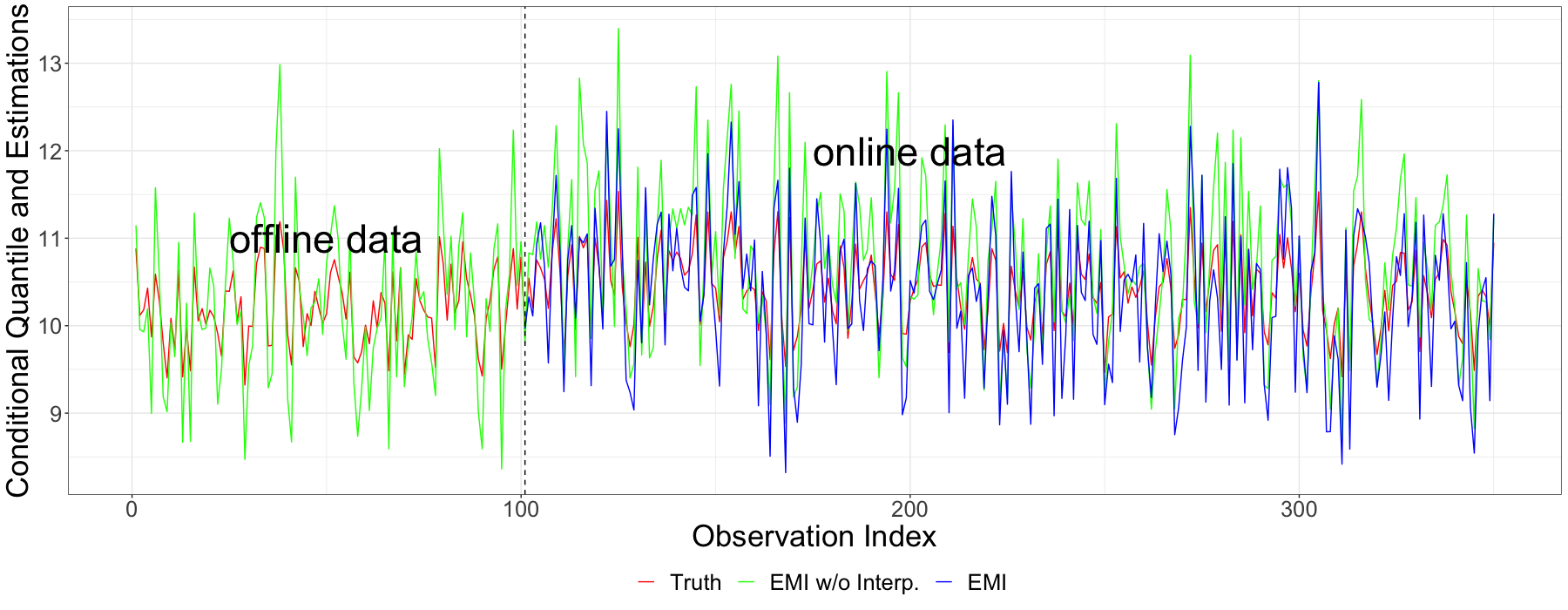}
		\centering
	}
	
	\subfigure[$\tau_N = 0.999$]{
		\includegraphics[scale = 0.23]{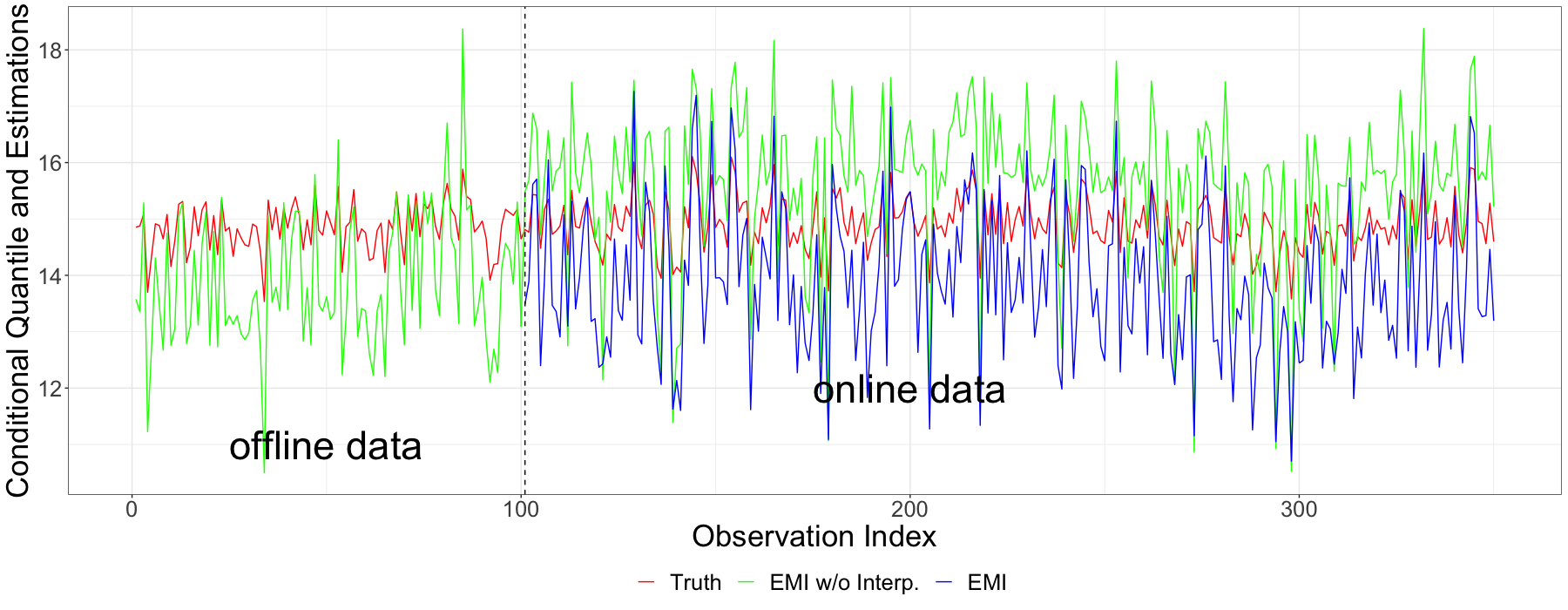}
		\centering
	}
	\caption{Illustrations of online scenarios. We set $N^{(\text{on})}=1000$ and $p=10$.}
		\label{fig:9}
\end{figure*}

\begin{figure*}[t!]
	\centering
	\subfigure[$\tau_N = 0.99$]{
		\includegraphics[scale = 0.23]{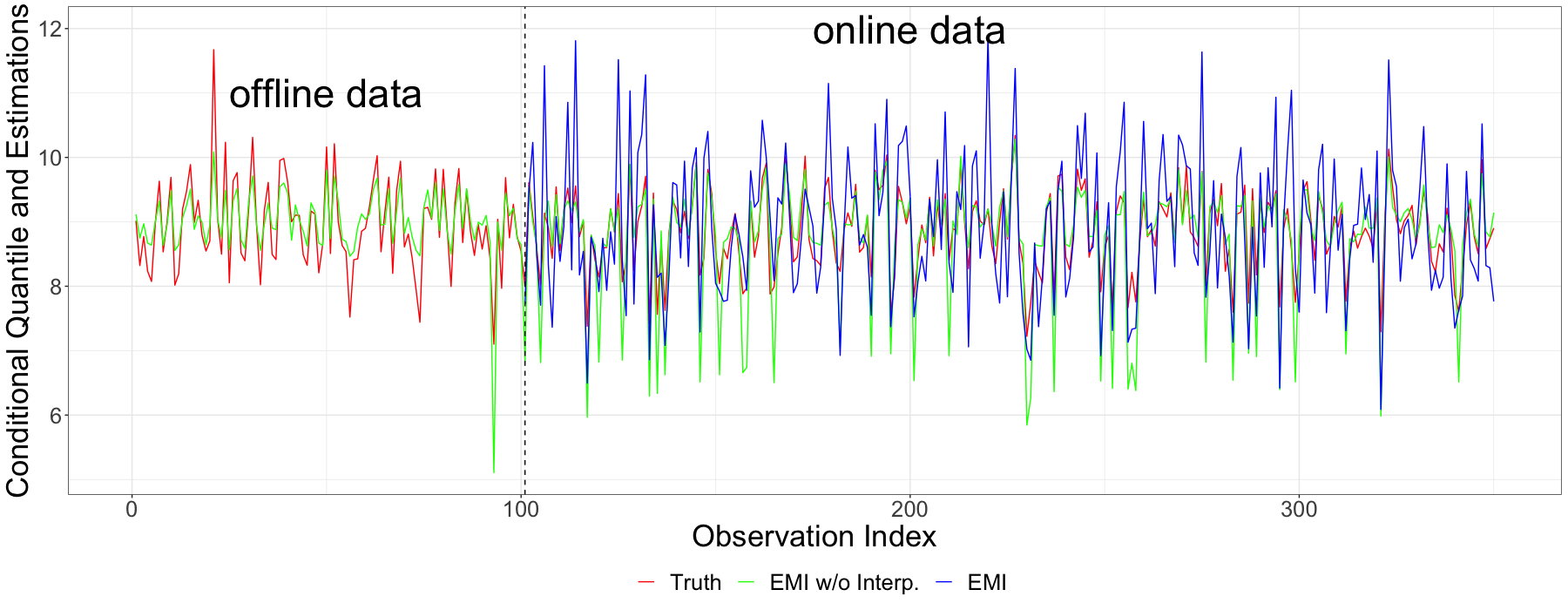}
		\centering
	}
	
	\subfigure[$\tau_N = 0.995$]{
		\includegraphics[scale = 0.23]{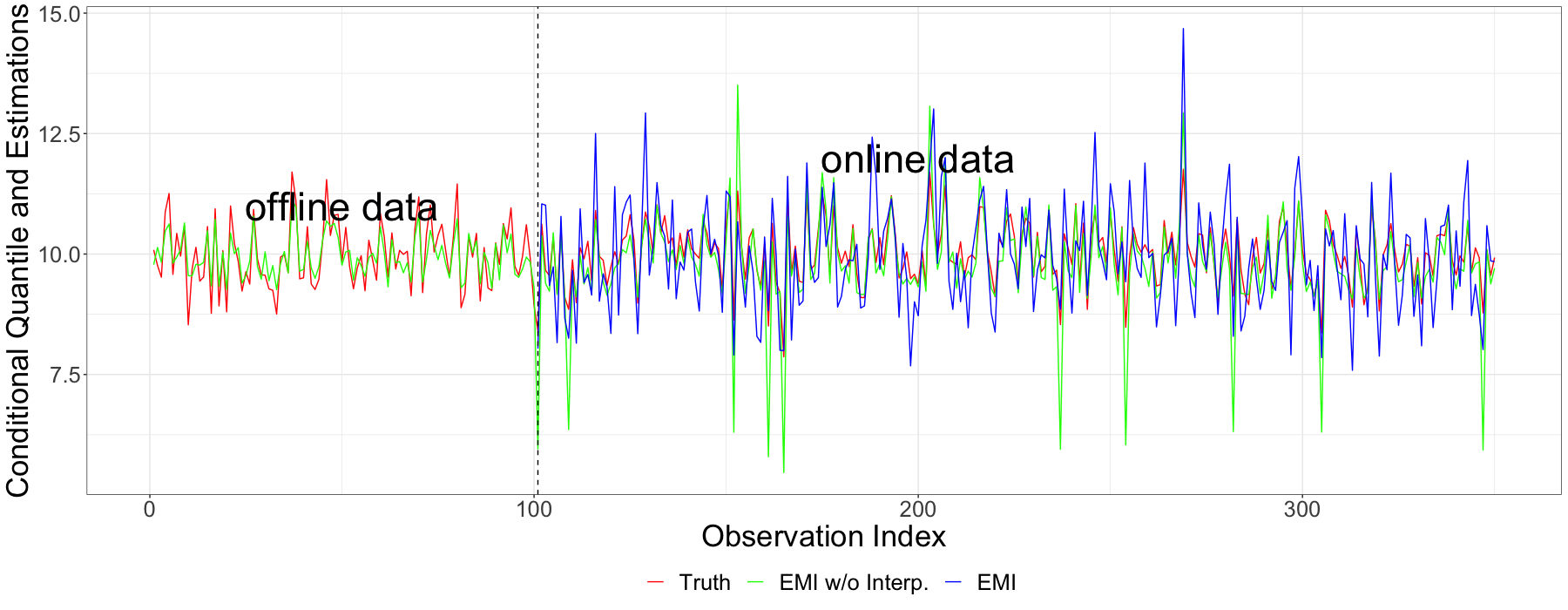}
		\centering
	}
	
	\subfigure[$\tau_N = 0.999$]{
		\includegraphics[scale = 0.23]{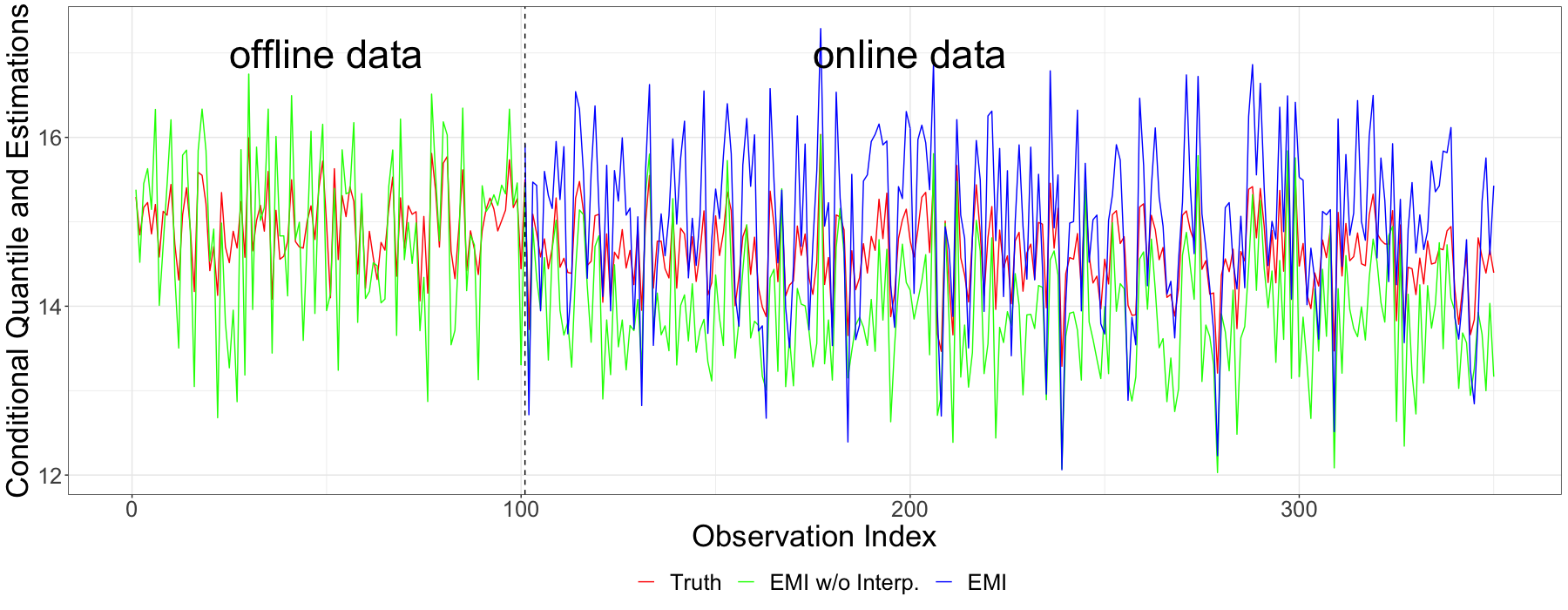}
		\centering
	}
	\caption{Illustrations of online scenarios. We set $N^{(\text{on})}=1000$ and $p=20$.}
		\label{fig:10}
\end{figure*}

\subsection{Results of Online Data}\label{sec3.4}
In our final scenario, we consider a more realistic situation where online data is observed incrementally, one data point at a time. At each time step $t$, we observe a new pair $(y_t^{(\text{on})}, \bx_t^{(\text{on})})$. The current observation set at time $t$, denoted as $(\mathcal{Y}^{(t)}, \mathcal{X}^{(t)})$, are the union of the new observation $(y_t^{(\text{on})}, \bx_t^{(\text{on})})$ and the observation set from the previous time step $t-1$. This is represented as:
\begin{equation*}
	\begin{split}
		\mathcal{Y}^{(t)} &= \mathcal{Y}^{(t-1)} \cup \{y_t^{(\text{on})}\}\\ \mathcal{X}^{(t)} &= \mathcal{X}^{(t-1)} \cup \{\bx_t^{(\text{on})}\},
	\end{split}
\end{equation*}
for $t=1,\ldots,T$. The initial observation set at time $t=0$ consists of only offline data as $\mathcal{Y}^{(0)} = \left\{y_i^{(\text{off})}\right\}$ and $\mathcal{X}^{(0)} = \left\{x_i^{(\text{off})}\right\}$. For comparison, the \emph{Linear} method solves the following linear quantile regression problem at each time step $t$:
\begin{equation*}
	\min _{\alpha,\bbeta} \sum_{(y,\bx)\in (\mathcal{Y}^{(t)}, \mathcal{X}^{(t)})}^{}\rho_{\tau_N}(y-\alpha-\bbeta^T\bx)
\end{equation*}
On the other hand, the \emph{EMI w/o. Interp.} method applies extrapolation based on the GPD and solves the following optimization problem at each time step $t$:
\begin{equation}\label{eq.222}
	\begin{split}
		\min _{\gamma,\sigma}\quad & -\sum_{z_i \in \mathcal{Z}^{(t)}}l(\gamma,\sigma|z_i)\\
		 s.t.\quad& \min _{\alpha,\bbeta} \sum_{(y,\bx)\in (\mathcal{Y}^{(t)}, \mathcal{X}^{(t)})}^{}\rho_{\tau_0}(y-\alpha-\bbeta^T\bx),
	\end{split}
\end{equation}
Here, $\mathcal{Z}^{(t)}$ is a set of transformed observations defined as $z_i=\left(y-\widetilde{\alpha}''(\tau_0)- \widetilde{\bbeta}''^\top(\tau_0)\bx_t^{(\text{on})} \right)_{+}$ for each $y \in \mathcal{Y}^{(t)}$, and $\widetilde{\alpha}''(\tau_0)$ and $\widetilde{\bbeta}''(\tau_0)$ are solutions of~\eqref{eq.222}. We put the results in Fig~\ref{fig:7a}, Fig~\ref{fig:7b} and Fig~\ref{fig:8}. Comparing the EMI method with the \emph{Linear} method, we observe that the performance of EMI is consistently better. Additionally, when compared to \emph{EMI w/o. Interp.}, the performance of EMI is slightly worse, which is consistent with previous experimental results. In Fig~\ref{fig:9} and Fig~\ref{fig:10}, we present visualizations of extreme conditional quantile estimates for online data streams. To keep the presentation concise, we select only a portion of the data. These figures demonstrate EMI's effectiveness in capturing changes in conditional quantiles.

\section{Application}
\begin{table*}[t!]
\centering
\caption{Summary statistics of covariates $\bX$.}
\label{table1}
	\begin{tabular}{l|cccccc}
\hline
\hline
$\bX$ & Name & Mean & Standard deviation & Skewness & Min. & Max. \\
\hline$x_1$& Market return & 0.16 & 2.23 & -0.29 & -15.35 & 12.00 \\
$x_2$& Three-month yield change & -0.22 & 24.43 & -0.57 & -182.00 & 192.00 \\
$x_3$ &Equity volatility & 0.83 & 0.43 & 3.06 & 0.28 & 4.70 \\
$x_4$& Credit spread change & -0.02 & 8.86 & 0.74 & -48.00 & 60.00 \\
$x_5$& Term spread change & 0.15 & 20.66 & 0.09 & -168.00 & 146.00 \\
$x_6$& TED spread & 121.52 & 94.42 & 1.74 & 6.34 & 591.00 \\
$x_7$& Real estate excess return & -0.08 & 2.38 & 0.04 & -14.49 & 16.58 \\
\hline
\end{tabular}
\end{table*}

\begin{table*}[t!]
	\centering
	\caption{Summary statistics of $Y$.}
	\label{table2}
	\begin{tabular}{l|cccc}
		\hline \hline
		$Y$ & Company & Start time & End time \\
		\hline
		S\&P & S\&P 500 & January 1st, 1971 & December 28th, 2002 \\
		AAPL & Apple Inc. & January 1st, 1981 & December 28th, 2002 \\
		BA & The Boeing Company & January 1st, 1971 & December 28th, 2002 \\
		JPM & JPMorgan Chase \& Co. & January 1st, 1981 & December 28th, 2002 \\
		\hline
	\end{tabular}
\end{table*}

In this section, we apply our algorithm EMI to estimate the extreme conditional quantile of a real dataset, consisting of the daily stock price of the S\&P 500 and three companies (see Table~\ref{table2}). The weekly return is derived from daily index observations. The covariates employed in our analysis include {\em ret} (weekly return), {\em marketret} (weekly market return), {\em yield3m} (three-month yield change), {\em mktsd} (equity volatility), {\em credit} (credit spread change), {\em term} (term spread change), {\em ted} (short-term TED spread), and {\em housing} (real estate excess return), as documented in (~\citeauthor{adrian2016covar}, \citeyear{adrian2016covar}). More specificity, the variables are defined as:
\begin{itemize}
	\item $Y$: The weekly loss, namely the weekly return;
	\item $x_1$: The weekly market return of S\&P 500;
	\item $x_2$: The change in the three-month yield from the Federal Reserve Board's $H.15$ release. We use the change in the three-month treasury bill rate because we find that the change, not the level, is most significant in explaining the tails of financial sector market-valued asset returns;
	\item $x_3$: Equity volatility, which is computed as the $22$-day rolling standard deviation of the daily CRSP equity market return;
    \item $x_4$: The change in the credit spread between Moody's Baarated bonds and the ten-year Treasury rate from the Federal Reserve Board's $H .15$ release;
	\item $x_5$: The change in the slope of the yield curve, measured by the spread between the composite long-term bond yield and the three-month bill rate obtained from the Federal Reserve Board's $H .15$ release;
	\item $x_6$: A short-term TED spread, which is defined as the difference between the three-month LIBOR rate and the three-month secondary market treasury bill rate. This spread measures short-term funding liquidity risk. We use the three-month LIBOR rate that is available from the British Bankers' Association, and obtain the three-month Treasury rate from the Federal Reserve Bank of New York;
	\item $x_7$:  The weekly real estate sector returns over the market financial sector return (from the real estate companies with SIC code $65$-$66$).
\end{itemize}

\begin{figure*}[t!]
	\centering
	\subfigure[S\&P.]{
	\begin{minipage}[t]{0.45\linewidth}
		\includegraphics[scale = 0.4]{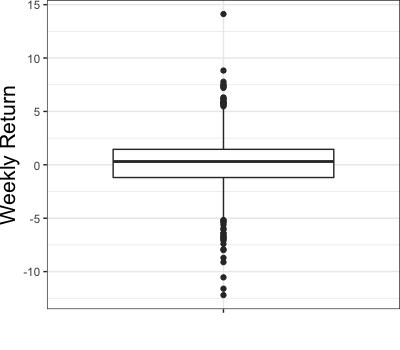}
		\centering
	\end{minipage}
	}
	\subfigure[AAPL.]{
	\begin{minipage}[t]{0.45\linewidth}
		\includegraphics[scale = 0.4]{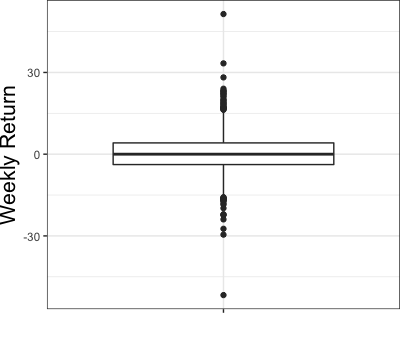}
		\centering
	\end{minipage}
	}
	
	\subfigure[BA.]{
	\begin{minipage}[t]{0.45\linewidth}
		\includegraphics[scale = 0.4]{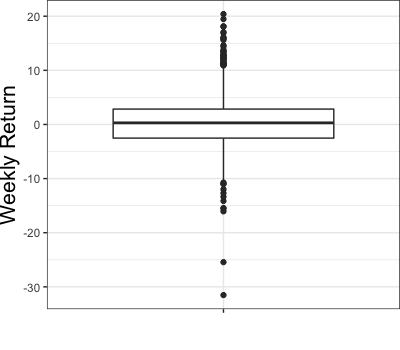}
		\centering
	\end{minipage}
	}
	\subfigure[JPM.]{
	\begin{minipage}[t]{0.45\linewidth}
		\includegraphics[scale = 0.4]{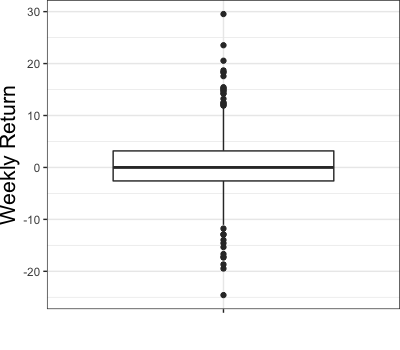}
		\centering
	\end{minipage}
	}
	\caption{Summary of the weekly return.}
	\label{fig:11}
\end{figure*}

\begin{figure*}[t!]
	\centering
		\includegraphics[scale = 0.22]{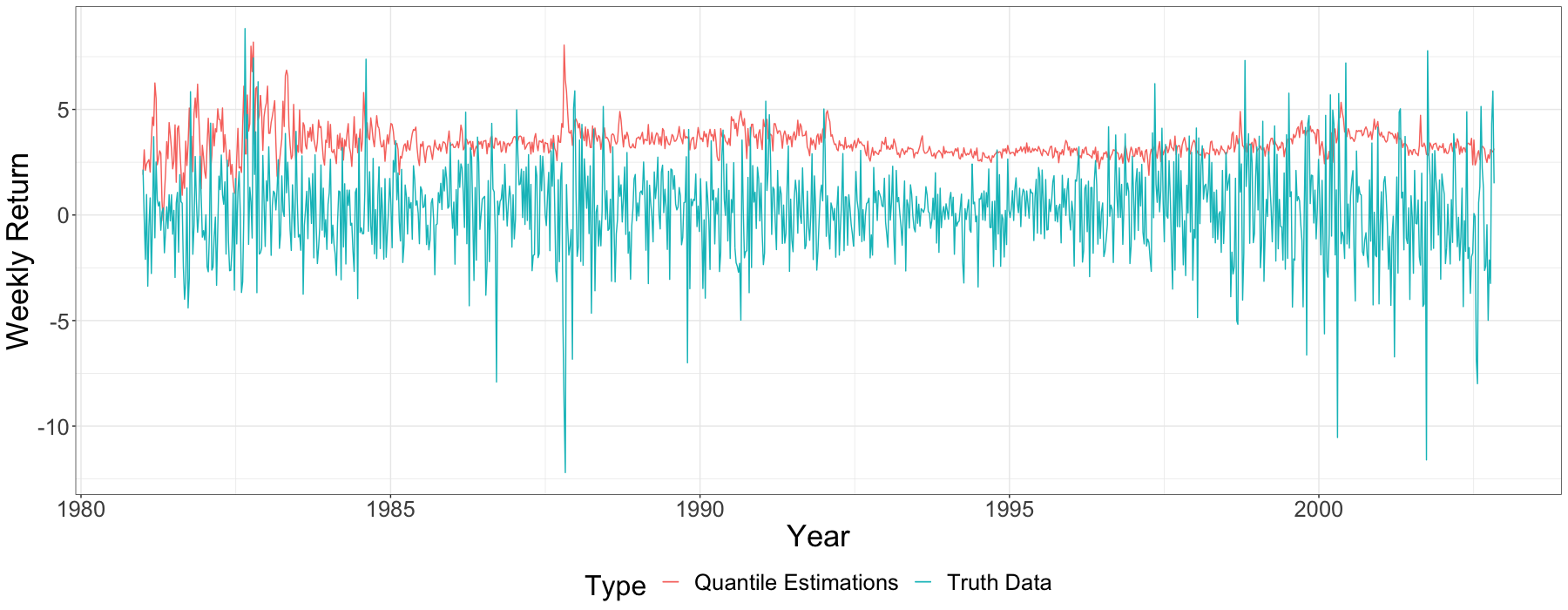}
		\centering
	\caption{Results for S\&P. The green curve is the estimated conditional quantile, and the blue curve is the truth response.}
	\label{fig:12}
\end{figure*}

We present a summary of the variables in Table~\ref{table1}. At time $t$, we assume that the response is related to covariates at time $t-1$. As a result, the quantile of response $Y_t$ depends on the covariate $\bX_{t-1}$ in a linear form:
\begin{equation*}
Q_{Y_t}(\tau|\bx)=\alpha(\tau)+ \bbeta^T(\tau)\bX_{t-1}.
\end{equation*}
In our data analysis, our primary focus is on estimating the extreme conditional quantile at the level $\tau=0.95$ for S\&P 500 and $\tau=0.99$ for three companies using EMI. Initially, we conduct a brief visual assessment to ascertain the suitability of the heavy-tail assumption for our data. As depicted in Fig.~\ref{fig:11}, the returns exhibit heavy tails on both ends. However, our emphasis lies in scrutinizing significant losses, directing our analysis towards the upper tail losses of the weekly returns, specifically, the extremely large losses. For S\&P 500 and each company, we follow the setting in Section~\ref{sec3.4} to estimate the extreme conditional quantile of future weekly returns employing historical data. To be specific, let $\mathcal{Y} = \{y_i\}_{i=1}^T$ and $\mathcal{X} = \{\bx_i\}_{i=1}^T$, where $T$ represents the total time span and $(y_i, \bx_i)$ denotes the $i$-th weekly data. At each time point $t = 521, 525, \ldots, T$, the observation set is defined as:
\begin{equation*}
	\mathcal{Y}^{(t)} = \{y_i\}_{i=t-520}^t,\quad \mathcal{X}^{(t)} = \{\bx_i\}_{i=t-520}^t.
\end{equation*}
In essence, we leverage the data from the past ten years to estimate the extreme conditional quantile of the subsequent four weekly returns.

\begin{figure*}[t!]
	\centering
		\includegraphics[scale = 0.22]{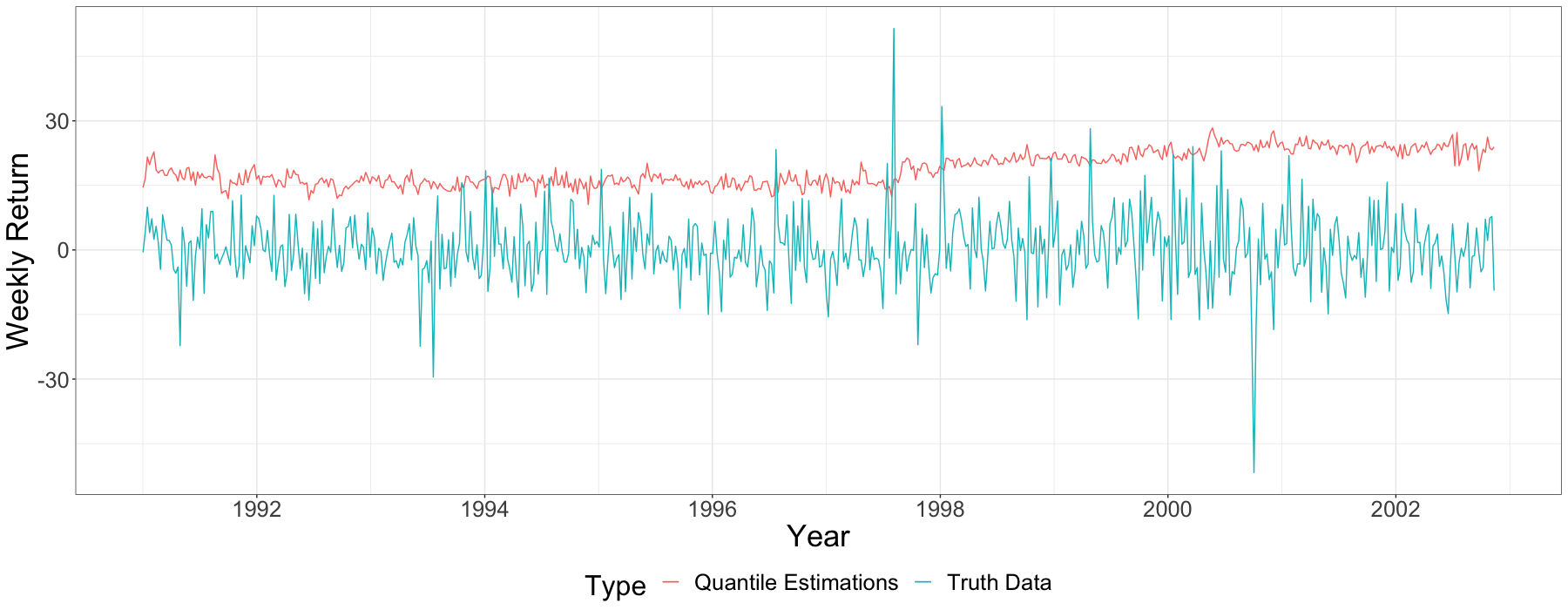}
		\centering
	\caption{Results for AAPL. The green curve is the estimated conditional quantile, and the blue curve is the truth response.}
	\label{fig:13}
\end{figure*}

\begin{figure*}[t!]
		\includegraphics[scale = 0.22]{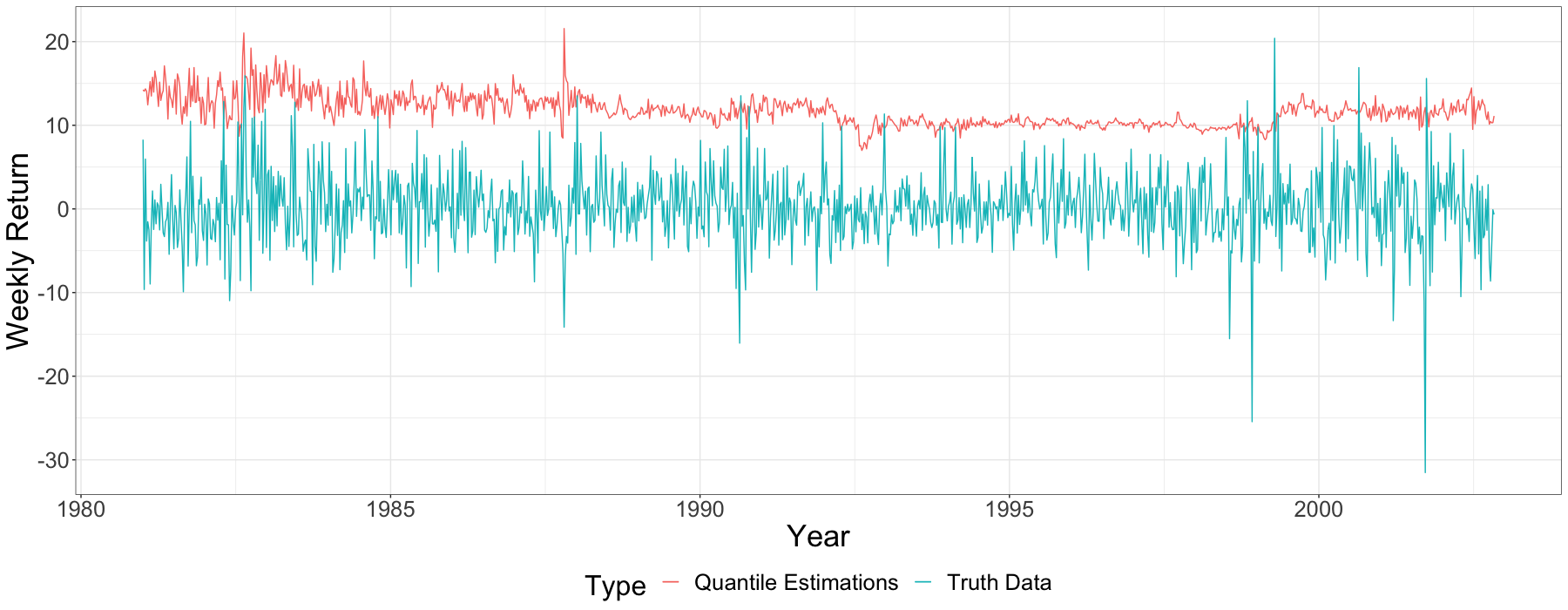}
		\centering
	\caption{Results for BA. The green curve is the estimated conditional quantile, and the blue curve is the truth response.}
	\label{fig:14}
\end{figure*}

\begin{figure*}[t!]
		\includegraphics[scale = 0.22]{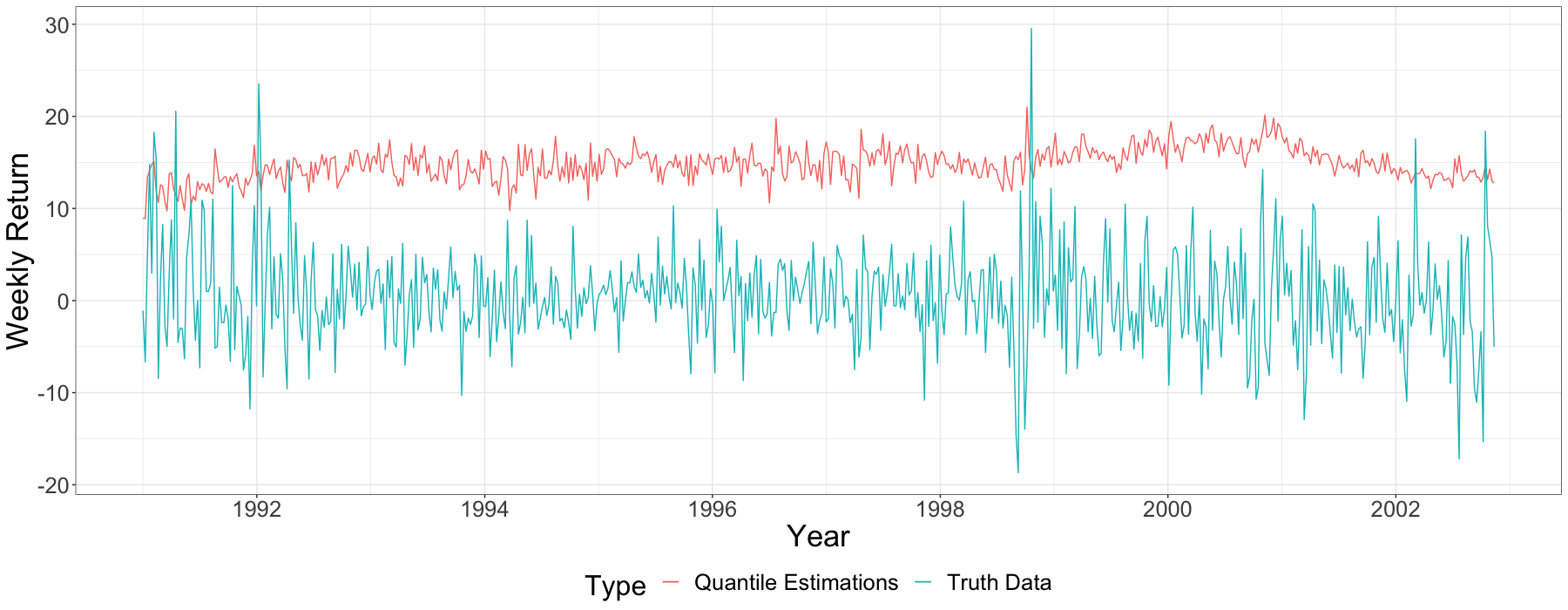}
		\centering
	\caption{Results for JPM. The green curve is the estimated conditional quantile, and the blue curve is the truth response.}
	\label{fig:15}
\end{figure*}

We summarize our results in Fig.~\ref{fig:12} (S\&P), Fig.~\ref{fig:13} (AAPL), Fig.~\ref{fig:14} (BA) and Fig.~\ref{fig:15} (JPM). 2008). We observe that the index of S\&P 500 exhibits the lowest conditional quantiles, indicating that it has the most robust capability to resist risk. Among the three companies studied, the conditional quantile of losses of AAPL is much larger compared to BA and JPM. It probably reflects the fact that technical companies may suffer more risk during the crisis period. In conclusion, our empirical analysis demonstrates that the proposed method for extreme conditional quantile serves as a good risk measure, particularly when considering loss values in tail regions.

\section{Conclusion}
In this paper, we introduce an algorithm called EMI designed for estimating extreme conditional quantiles. Operating with finite offline observations, our approach approximates the exceedance above a predetermined threshold using the GPD. We formulate the extrapolation process by rewriting it as bilevel programming, making it amenable to efficient solutions through conventional optimization techniques. To enhance flexibility and computational efficiency, we incorporate B-spline interpolation for covariate-dependent parameters. This interpolation capability empowers EMI to estimate conditional quantiles for potentially infinite online data, avoiding the computational complexity associated with GPD fitting. Empirical experiments convincingly demonstrate that our proposed EMI consistently outperforms the linear conditional regression model.

\section{Acknowledgment}	
Yanxi Hou's work was supported by the MOE Laboratory for National Development and Intelligent Governance, Fudan University, the National Natural Science Foundation of China Grant 72171055, and the Natural Science Foundation of Shanghai Grant 20ZR1403900.

\bibliography{EMI_ref.bib}

\begin{thebibliography}{28}
\providecommand{\natexlab}[1]{#1}
\providecommand{\url}[1]{{#1}}
\providecommand{\urlprefix}{URL }
\providecommand{\doi}[1]{\url{https://doi.org/#1}}
\providecommand{\eprint}[2][]{\url{#2}}
 \bibcommenthead

\bibitem[{Adrian and Brunnermeier(2016)}]{adrian2016covar}
Adrian T, Brunnermeier MK (2016) Covar. American Economic Review
  106(7):1705--1741

\bibitem[{Aiyoshi and Shimizu(1981)}]{aiyoshi1981hierarchical}
Aiyoshi E, Shimizu K (1981) Hierarchical decentralized systems and its new
  solution by a barrier method. IEEE Transactions on Systems, Man and
  Cybernetics 11(6):444--449

\bibitem[{Allen et~al(2012)Allen, Bali, and Tang}]{allen2012does}
Allen L, Bali TG, Tang Y (2012) Does systemic risk in the financial sector
  predict future economic downturns? The Review of Financial Studies
  25(10):3000--3036

\bibitem[{Angelo et~al(2013)Angelo, Krempser, and
  Barbosa}]{angelo2013differential}
Angelo JS, Krempser E, Barbosa HJ (2013) Differential evolution for bilevel
  programming. In: 2013 IEEE Congress on Evolutionary Computation, pp 470--477

\bibitem[{Balkema and De~Haan(1974)}]{balkema1974residual}
Balkema AA, De~Haan L (1974) Residual life time at great age. The Annals of
  Probability 2(5):792--804

\bibitem[{Drees et~al(2004)Drees, Ferreira, and De~Haan}]{drees2004maximum}
Drees H, Ferreira A, De~Haan L (2004) On maximum likelihood estimation of the
  extreme value index. Annals of Applied Probability 14(3):1179--1201

\bibitem[{Ferreira and De~Haan(2015)}]{ferreira2015block}
Ferreira A, De~Haan L (2015) On the block maxima method in extreme value
  theory: {PWM} estimators. The Annals of Statistics 3(1):276--298

\bibitem[{Gaber et~al(2005)Gaber, Zaslavsky, and
  Krishnaswamy}]{gaber2005mining}
Gaber MM, Zaslavsky A, Krishnaswamy S (2005) Mining data streams: {A} review.
  ACM Sigmod Record 34(2):18--26

\bibitem[{Granello and Wheaton(2004)}]{granello2004online}
Granello DH, Wheaton JE (2004) Online data collection: Strategies for research.
  Journal of Counseling \& Development 82(4):387--393

\bibitem[{He et~al(2022)He, Peng, Zhang, and Zhao}]{he2022risk}
He Y, Peng L, Zhang D, et~al (2022) Risk analysis via generalized pareto
  distributions. Journal of Business \& Economic Statistics 40(2):852--867

\bibitem[{Hosking et~al(1985)Hosking, Wallis, and Wood}]{hosking1985estimation}
Hosking JRM, Wallis JR, Wood EF (1985) Estimation of the generalized
  extreme-value distribution by the method of probability-weighted moments.
  Technometrics 27(3):251--261

\bibitem[{Hou et~al(2022)Hou, Kang, Lo, and Peng}]{hou2022three}
Hou Y, Kang SK, Lo CC, et~al (2022) Three-step risk inference in insurance
  ratemaking. Insurance: Mathematics and Economics 105:1--13

\bibitem[{Koenker and Bassett~Jr(1978)}]{koenker1978regression}
Koenker R, Bassett~Jr G (1978) Regression quantiles. Econometrica 46(1):33--50

\bibitem[{Koenker and Hallock(2001)}]{koenker2001quantile}
Koenker R, Hallock KF (2001) Quantile regression. Journal of Economic
  Perspectives 15(4):143--156

\bibitem[{Li and Wang(2019)}]{li2019extreme}
Li D, Wang HJ (2019) Extreme quantile estimation for autoregressive models.
  Journal of Business \& Economic Statistics 37(4):661--670

\bibitem[{Luo et~al(1996)Luo, Pang, and Ralph}]{luo1996mathematical}
Luo ZQ, Pang JS, Ralph D (1996) Mathematical Programs with Equilibrium
  Constraints. Cambridge University Press, Cambridge, England

\bibitem[{Lv et~al(2007)Lv, Hu, Wang, and Wan}]{lv2007penalty}
Lv Y, Hu T, Wang G, et~al (2007) A penalty function method based on
  {K}uhn--{T}ucker condition for solving linear bilevel programming. Applied
  Mathematics and Computation 188(1):808--813

\bibitem[{Mathieu et~al(1994)Mathieu, Pittard, and
  Anandalingam}]{mathieu1994genetic}
Mathieu R, Pittard L, Anandalingam G (1994) Genetic algorithm based approach to
  bi-level linear programming. Operations Research 28(1):1--21

\bibitem[{Naveau et~al(2009)Naveau, Guillou, Cooley, and
  Diebolt}]{naveau2009modelling}
Naveau P, Guillou A, Cooley D, et~al (2009) Modelling pairwise dependence of
  maxima in space. Biometrika 96(1):1--17

\bibitem[{Pickands~III(1975)}]{pickands1975statistical}
Pickands~III J (1975) Statistical inference using extreme order statistics. The
  Annals of Statistics 3(1):119--131

\bibitem[{Resnick(2008)}]{resnick2008extreme}
Resnick SI (2008) Extreme Values, Regular Variation, and Point Processes.
  Springer Science \& Business Media, Berlin, Germany

\bibitem[{Sinha et~al(2014)Sinha, Malo, Frantsev, and Deb}]{sinha2014finding}
Sinha A, Malo P, Frantsev A, et~al (2014) Finding optimal strategies in a
  multi-period multi-leader--follower {S}tackelberg game using an evolutionary
  algorithm. Computers \& Operations Research 41:374--385

\bibitem[{Sinha et~al(2017)Sinha, Malo, and Deb}]{sinha2017review}
Sinha A, Malo P, Deb K (2017) A review on bilevel optimization: From classical
  to evolutionary approaches and applications. IEEE Transactions on
  Evolutionary Computation 22(2):276--295

\bibitem[{Smith(1987)}]{smith1987estimating}
Smith RL (1987) Estimating tails of probability distributions. The Annals of
  Statistics 15(3):1174--1207

\bibitem[{Velthoen et~al(2019)Velthoen, Cai, Jongbloed, and
  Schmeits}]{velthoen2019improving}
Velthoen J, Cai JJ, Jongbloed G, et~al (2019) Improving precipitation forecasts
  using extreme quantile regression. Extremes 22:599--622

\bibitem[{Wang and Li(2013)}]{wang2013estimation}
Wang HJ, Li D (2013) Estimation of extreme conditional quantiles through power
  transformation. Journal of the American Statistical Association
  108(503):1062--1074

\bibitem[{Wang et~al(2012)Wang, Li, and He}]{wang2012estimation}
Wang HJ, Li D, He X (2012) Estimation of high conditional quantiles for
  heavy-tailed distributions. Journal of the American Statistical Association
  107(500):1453--1464

\bibitem[{Xu et~al(2022)Xu, Hou, and Li}]{xu2022prediction}
Xu W, Hou Y, Li D (2022) Prediction of extremal expectile based on regression
  models with heteroscedastic extremes. Journal of Business \& Economic
  Statistics 40(2):522--536

\end{thebibliography}

\end{document}